\documentclass[aip,amsmath,amssymb,reprint,floatfix,pof]{revtex4-1}
\usepackage{graphicx}% Include figure files
\usepackage{dcolumn}% Align table columns on decimal point
\usepackage{bm}% bold math
\usepackage[utf8]{inputenc}
\usepackage[T1]{fontenc}
\usepackage{mathptmx}
\usepackage{etoolbox}
\usepackage{verbatim}
\usepackage{xcolor}
\usepackage{soul}
\usepackage{orcidlink}
\usepackage{hyperref}% add hypertext capabilities
\hypersetup{colorlinks,%,%
 linkcolor=blue,%
 citecolor=blue,%
 urlcolor=blue
}

\DeclareMathOperator{\sech}{sech}

\begin{document}

\title{Dynamically Stable Vortices in Exciton-Polariton Condensates Engineered by Repulsive Interactions}

\author{Palanivel Raman}
\affiliation{Department of Physics, Bharathidasan University, Tiruchirappalli 620 024, Tamil Nadu, India}
\affiliation{Centre for Nonlinear Science (CeNSc), Government College for Women (Autonomous), Kumbakonam 612001, Tamil Nadu, India}

\author{Ramaswamy Radha\orcidlink{0000-0002-5969-580X}}
%\email{vittal.cnls@gmail.com}
\affiliation{Centre for Nonlinear Science (CeNSc), Government College for Women (Autonomous), Kumbakonam 612001, Tamil Nadu, India}

\author{Pankaj Kumar Mishra\orcidlink{0000-0003-1566-0012}}
%\email{pankaj.mishra@iitg.ac.in}
\affiliation{Department of Physics, Indian Institute of Technology Guwahati, Guwahati 781039, Assam, India}

\author{Paulsamy Muruganandam\orcidlink{0000-0002-3246-3566}}
%\email{anand@bdu.ac.in}
\affiliation{Department of Physics, Bharathidasan University, Tiruchirappalli 620 024, Tamil Nadu, India}
\affiliation{Department of Medical Physics, Bharathidasan University, Tiruchirappalli 620024, Tamil Nadu, India}

\date{\today}

\begin{abstract}
We present an analytical and numerical study of the dynamics and stability of exciton-polariton condensates described by the open-dissipative Gross-Pitaevskii equation, incorporating both binary and short-range three-body interactions. Using an asymptotic description, we identify the parameter regime and derive equations for the instability amplitude, providing insights into vortex formation via the snake instability of dark solitons. We find that a repulsive three-body interaction, when combined with a binary interaction, supports stable vortex-antivortex pair formation. On the other hand, the reinforcement of attractive three-body interactions with binary interaction triggers the emergence of snake instability, leading to boundary-driven vortex disintegration. The time evolution of the instability under the influence of reservoir effects indicates that the boundary effects are more pronounced, to the extent of destabilizing the vortices with attractive three-body interactions compared to repulsive three-body interactions, thereby underscoring the stable nature of vortices in the repulsive domain.
\end{abstract}

\maketitle

\section{Introduction}

Exciton-polariton condensates, which arise from the strong coupling between photons and excitons in semiconductor microcavities, represent a significant paradigm shift in the study of quantum fluids of light and matter~\cite{Kasprzak2006, balili2007bose, Deng2010, Assmann2011}. Owing to their hybrid light-matter nature, polaritons inherit an exceptionally small effective mass from their photonic component~\cite{Bao2019, Assmann2011}, enabling Bose‑Einstein condensation at relatively high temperatures~\cite{Kasprzak2006, byrnes2014exciton}, including room temperature in state of the art perovskite based microstructures~\cite{Bao2019, Peng2024}. This makes them a versatile platform for exploring macroscopic quantum phenomena beyond the constraints of ultracold atomic gases~\cite{Assmann2011, Carusotto2013}. However, exciton-polariton condensates are short-lived, and stabilizing them continues to be an arduous task even today~\cite{Deng2010,Sabari2022}.

A defining feature of exciton-polariton condensates is their intrinsically nonequilibrium character~\cite{Deng2010, PhysRevB.109.195432}. Unlike equilibrium atomic Bose-Einstein condensates, polariton condensates are sustained by continuous optical pumping and are subject to rapid radiative decay~\cite{Carusotto2013, Wouters2007, Keeling2008}. Despite this driven-dissipative character, they exhibit the hallmark  of quantum degeneracy, including superfluidity, long-range order, and the formation of quantized vortices~\cite{Amo2009, Lagoudakis2008, Keeling2008}. The interplay of nonlinearity, gain, and loss in these systems leads to a rich landscape of phenomena, such as pattern formation, spontaneous symmetry breaking, and topological excitations~\cite{dominici2015vortex, Flayac2010}. 
In particular, exciton-polariton condensates have emerged as an ideal platform for two-dimensional quantum hydrodynamics, establishing key signatures of quantum turbulence and scale-invariant energy cascades in fluids of light~\cite{Panico2023}.

This unique non-equilibrium physics has stimulated extensive experimental and theoretical research, leading to the development of polariton lasers~\cite{Christopoulos2007PolaritonLaser}, ultrafast optical switches~\cite{Amo2010Switching}, and robust nonlinear excitations such as dark solitons and vortices~\cite{Sich2011, Egorov2009, Smirnov2014}. Increasing attention has also been directed towards
many-body correlation effects, including the formation of composite polaronic quasiparticles mediated by biexciton and triexciton resonances~\cite{Parish2019Spectroscopic}. To date, however, the dominant theoretical treatment of interactions in polariton condensates has been largely restricted to effective two-body processes, resulting in cubic nonlinearities within the mean-field description.        

A natural, yet comparatively unexplored, extension of this framework involves higher-order interactions. In particular, effective three-body interactions can arise from virtual excitation processes, while excitonic saturation and phase-space filling at elevated densities can substantially modify the stability of the condensate and its ground-state phase structure~\cite{Levinsen2015, Hammond2022}. 
While such interactions have been studied extensively in equilibrium atomic Bose-Einstein condensates (BECs), where they are known to stabilize quantum droplets and suppress collapse~\cite{Sudharsan2013, Ramesh2010, sasireka2025stability}, their emergence and implications in non-equilibrium polariton systems, including the formation of quantum droplets of light, are beginning  to be explored~\cite{qbz5-df6g}. The specific role of three-body interactions in stabilizing topological excitations, such as quantized vortices and solitons, against dissipation and driving in polariton condensates constitutes a significant gap in the literature.
%While such interactions have been studied extensively in equilibrium atomic Bose-Einstein condensates (BECs), where they are known to stabilize quantum droplets and suppress collapse~\cite{Sudharsan2013, Ramesh2010, sasireka2025stability}, their implications in non-equilibrium polariton systems remain largely unexplored. The specific role of three-body interactions in stabilizing topological excitations, such as quantized vortices and solitons, against dissipation and driving in polariton condensates constitutes a significant gap in the literature. 
In this paper, we aim to address this gap by systematically investigating the impact of three-body interactions on the stability and dynamics of topological defects in exciton-polariton condensates.

In this work, we theoretically investigate the impact of binary and effective three-body interactions on the dynamical stability and non-equilibrium evolution of topological defects in driven-dissipative exciton-polariton condensates. By incorporating higher-order nonlinearities arising from phase-space filling, excitonic saturation, and multi-body polariton scattering beyond the low-density limit~\cite{SedovE, Carusotto2013} into an open-dissipative Gross--Pitaevskii model, we establish a comprehensive theoretical and numerical framework for studying dark-soliton and vortex dynamics. Through an asymptotic analysis of the transverse snake instability, we elucidate how the sign and strength of the three-body interaction govern the spatial growth of perturbations along dark-soliton centerlines, building upon established driven-dissipative soliton dynamics~\cite{Smirnov2014, Maitre2020, Claude2020}. Crucially, we demonstrate that repulsive three-body interactions ($\chi > 0$) robustly suppress the snake instability and prolong vortex lifetimes, whereas attractive interactions ($\chi < 0$) accelerate transverse decay. These results establish effective three-body interactions as a versatile control parameter for manipulating the stability and non-equilibrium dynamics of topological excitations in quantum fluids of light.

The remainder of this paper is organized as follows. In Sec.~\ref{sec:model}, we introduce the mean-field open-dissipative Gross--Pitaevskii framework coupled to an excitonic reservoir, perform the dimensionless scaling, and map the dimensionless parameters to experimental microcavity platforms. We also examine the homogeneous steady-state solutions and discuss the dark soliton background profiles under weak and strong pumping regimes. In Sec.~\ref{sec:AsympDark}, we formulate a multiscale asymptotic perturbation analysis to derive the analytical evolution equations governing the transverse snake instability and dark soliton dynamics. Section~\ref{sec:num} presents full two-dimensional numerical simulations of the governing equations, comparing the baseline binary interaction case with attractive and repulsive three-body regimes, and quantitatively characterizing vortex stability, phase topologies, and lifetimes. Finally, our principal conclusions and outlook for future experimental realizations are summarized in Sec.~\ref{sec:con}. In the Appendix~\ref{appendix:velocity_derivation}, we provide a detailed derivation of the hydrodynamics velocity field.

\section{The mean-field Model}
\label{sec:model}

The dynamics of an exciton-polariton condensate within the mean-field framework is described by a coupled driven-dissipative system, comprising the Gross-Pitaevskii (GP) equation for the condensate wavefunction and a rate equation for the exciton reservoir of the following form~\cite{Smirnov2014, Sabari2022, SedovE}:
 \begin{subequations}
\begin{align}
\mathrm{i} \frac{\partial \Psi}{\partial t} &= \left[ -\frac{\hbar^{2}}{2m} \nabla^{2} + U(\vec{r},t) 
%\right. \notag \\ &\quad \left. 
+ \frac{\mathrm{i}\hbar}{2} \left( R n_{R} - \gamma_{C} \right) \right] \Psi, \label{eq:psi_old} \\
\frac{\partial n_{R}}{\partial t} &= P(\vec{r},t) - \left( \gamma_{R} + R \left| \Psi \right|^{2} \right) n_{R}. \label{density_old}
\end{align}
\label{eq:pbec_model_old}
\end{subequations}
Here, the effective potential is given by $U(\vec{r},t)= g_{C}|\Psi|^{2}+ \chi |\Psi|^{4}+ g_{R}n_{R}+ g_{R}n_{R}^{2}$. The $g_{C}$ denotes the strength of two-body nonlinear interactions, $\chi $ represents the strength of three-body interactions, and $g_{R}$ accounts for the interaction between the reservoir and the polariton condensate. The parameter $R$ corresponds to the stimulated scattering rate, $\gamma_{C}$ is the loss rate of the condensate polariton, and $\gamma_{R}$ denotes the decay rate of the exciton reservoir given by Eq.~\eqref{density_old}. High-energy exciton-like particles are injected into the reservoir by laser pump $P(\vec{r},t)$. 
Under continuous wave and spatially uniform pumping, where $P(\vec{r},t) = P_{0} = \mbox{constant}$, the steady-state solution is given by
\begin{align}
\Psi(\vec{r},t)= \Psi_{0} &= \sqrt{n_{C}^{0}}\mathrm{e}^{-\mathrm{i}(E_{0}/\hbar)t}, \quad n_{R}(\vec{r},t) = n_{R}^{0}.
\label{eq:cw-epc}
\end{align}
where, $n_{C}^{0}$ and $n_{R}^{0}$ denote the steady state densities corresponding to the condensate polariton and reservoir, respectively. In the weak pumping limit, the condensate density vanishes, i.e. $n_{C}^{0}=0$, while the reservoir density reaches a steady state density $n_{R}^{0}= P_{0}/ \gamma_{R}$ with the balance gain and loss at the threshold pump power. When the pumping exceeds this threshold as defined by Eq.~\eqref{eq:cw-epc}, the condensate density becomes non-zero, i.e., $n_{C}^{0}$, and a macroscopic condensate emerges. Above threshold, the steady-state homogeneous densities of the condensate and reservoir are given by:
\begin{align}
n_{C}^{0} & = (P_{0}- P_{\mbox{th}})/ \gamma_{C}, \quad n_{R}^{0}= n_{R}^{\mbox{th}}= \gamma_{C}/R,
\end{align}
The model can be written in dimensionless form by scaling the unit of healing length $r_{h}= \hbar/(Mc_{s})$ and time as $\tau_{0}=r_{h}/c_{s}$. Here, $c_{s}=(g_{C}n_{C}^{*}/M)^{1/2}$ is a local sound velocity in the condensate and $n_{C}^{*}$ is the characteristic density of the condensate. In dimensionless form, the wavefunction $\bar{\Psi}= \mathrm{\Psi}/\sqrt{n_{C}^{*}}$ and the reservoir density are given by $\bar n_{R}=n_{R}/n_{C}^{*}$. Under these circumstances, Eq.~\eqref{eq:pbec_model_old} becomes,
\begin{subequations}
\begin{align}
\mathrm{i} \frac{\partial \bar {\Psi}}{\partial t} & = - \frac{1}{2}\nabla^{2} \bar{\Psi}+ \bar{U}(\vec{r},t)+ \frac{\mathrm{i}}{2} \left(\bar{R} \bar n_{R}- \bar{\gamma}_{C} \right) \bar{\Psi}, \label{psi_modified} \\
\frac{\partial \bar{n}_{R}}{\partial t} &= \bar{P} - (\bar{\gamma}_{R} \bar n_{R} + \bar{R} \left| \bar{\Psi} \right|^{2} \bar n_{R}). \label{eq:rate0}
\end{align}
\label{eq:model}
\end{subequations}
Here, $\bar{U}(\vec{r},t)= |\bar{\Psi}|^{2}+ \bar\chi  |\bar{\Psi}|^{4} + \bar{g}_{R} \bar{n}_{R}+ \bar{g}_{R} \bar{n}_{R}^{2}$. The other dimensionless parameters are 
\begin{align*}
& \bar{\chi}= \frac{\chi n_{C}^{*}}{g_{C}}, \quad \bar{g}_{R}= \frac{g_{R}}{g_{C}}, \quad \bar{\gamma}_{C}= \frac{\hbar \gamma_{C}}{g_{C}n_{C}^{*}}, \quad \bar{\gamma}_{R}= \frac{\hbar \gamma_{R}}{g_{C}n_{C}^{*}}, \\&  \bar{R}= \frac{\hbar R}{g_{C}}, \quad  \bar{P}= \frac{\hbar P}{g_{C}(n_{C}^{*})^{2}}.
\end{align*}
For continuous background density, the characteristic value of the condensate density is given by $n_{C}^{*}= n_{C}^{0}$. Using this relation, the homogeneous steady state solution can be represented as \cite{Smirnov2014}
\begin{align}
 \bar{\Psi}_{0} = \mbox{exp}(-i \bar{\omega}t), \quad \bar{\omega} = \bar{g}_{R}n_{R}^{th}(1+n_{R}^{th}).
\end{align}
Let's consider the perturbations of the condensate wavefunction and the reservoir density in the general form as \cite{Ostrovskaya2012DissipativeSolitons}
\begin{align}
\bar{\Psi} = \bar{\Psi}_{0}(t)\psi(\vec{r},t), \quad \bar{n}_{R}= \bar{n}_{R}^{th}+ m_{R}(\vec{r},t).
\end{align}
In the homogeneous steady state, the condensate density is normalized as $|\bar{\Psi}_{0}|^{2}=1$, and gain--loss balance requires $\bar{R} \bar{n}_{R}^{\text{th}} = \bar{\gamma}_{C}$. Under these conditions, the homogeneous condensate evolves as
\begin{equation}
    \bar{\Psi}_{0}(t) = \mathrm{e}^{-\mathrm{i}\mu t},
\end{equation}
where the dimensionless chemical potential is given by $\mu = 1 + \bar{\chi} + \bar{g}_{R}\bar{n}_{R}^{\text{th}}$. Applying the rotating-frame transformation $\bar{\Psi}(\mathbf{r},t) = \mathrm{e}^{-\mathrm{i}\mu t}\psi(\mathbf{r},t)$, the driven--dissipative Gross--Pitaevskii equation gets transformed to
\begin{align}
    \mathrm{i} \frac{\partial \psi}{\partial t} = & \left[ -\frac{1}{2} \nabla^{2} + (|\psi|^{2}-1) + \bar{\chi}(|\psi|^{4}-1) + \bar{g}_{R} m_{R} \notag \right. \\ & \left. + \frac{\mathrm{i}}{2}\bar{R} m_{R} \right]\psi,
\end{align}
where $m_{R} = \bar{n}_{R} - \bar{n}_{R}^{\text{th}}$ represents the local reservoir perturbation relative to its steady-state threshold value $\bar{n}_{R}^{\text{th}}$. Here, the uniform phase factor $\mathrm{e}^{-\mathrm{i}\mu t}$ shifts the energy reference by the constant interaction energy $\mu = 1 + \bar{\chi} + \bar{g}_{R} \bar{n}_{R}^{\text{th}}$, removing the background phase evolution without affecting the condensate density, phase gradients, or superfluid current.

When we substitute the above in equations (\ref{psi_modified}) and ~(\ref{eq:rate0}), we obtain
\begin{subequations}\label{eq:model_dimless}
\begin{align}
\mathrm{i} \frac{\partial \psi}{\partial t} = & \left[ -\frac{1}{2} \nabla^{2} -(1-|\psi|^{2})-\bar\chi(1-|\psi|^{4})+ \bar{g}_{R}m_{R} \right.  \notag \\  &  \left. 
+ \bar{g}_{R}m_{R}^{2} + \frac{i}{2}\bar{R}m_{R}\right]\psi, \label{eq:Psi2}\\
 \frac{\partial m_{R}}{\partial t} = & \, \bar{\gamma}_{C}(1-|\psi|^{2})- \bar{\gamma}_{R}m_{R}- \bar{R}|\psi|^{2}m_{R}, \label{eq:rate}
\end{align} 
\end{subequations}
where $\bar{\gamma}_{R} = \bar{R}/(P_{0}/P_{th}-1)$.

\subsection{Experimental Relevance and Parameter Mapping}

To establish the experimental relevance of our theoretical framework, we map the dimensionless parameters of the open-dissipative Gross--Pitaevskii model onto physical scales characteristic of semiconductor microcavity platforms~\cite{Carusotto2013, Wouters2007}. These include cryogenic GaAs quantum wells and CdTe microcavities, as well as room-temperature platforms based on lead-halide perovskites, such as CsPbBr$_3$, and monolayer transition-metal dichalcogenides (TMDs), including MoS$_2$ and WS$2$~\cite{Carusotto2013, Estrecho2019}. For typical GaAs and CdTe microcavities, the polariton effective mass is of order $m^* \sim 10^{-4}m_e$, where $m_e$ is the free-electron mass. Taking $x_0=1~\mu\mathrm{m}$ and $t_0=1~\mathrm{ps}$ as characteristic length and time scales, respectively, gives an energy unit $\hbar/t_0\simeq0.658~\mathrm{meV}$. Typical polariton decay rates are $\gamma_C\sim0.05$--$0.5~\mathrm{ps}^{-1}$, corresponding to lifetimes of a few to several tens of picoseconds, while reservoir relaxation rates may span $\gamma_R\sim0.001$--$0.05~\mathrm{ps}^{-1}$~\cite{Wouters2007, Bobrovska2014}. The stimulated scattering coefficient $R$ is typically of the order $10^{-3}$--$10^{-1}~\mathrm{ps}^{-1}\mu\mathrm{m}^{2}$, yielding condensation thresholds $P_{\mathrm{th}}=\gamma_C\gamma_R/R$ within experimentally accessible pumping regimes~\cite{Wouters2007}. The polariton-polariton interaction strength is typically $g_C\sim0.5$--$5~\mu\mathrm{eV}\,\mu\mathrm{m}^{2}$, while the reservoir-induced interaction is commonly taken to be of comparable magnitude, with $g_R$ of order $g_C$~\cite{Estrecho2019}.

The higher-order nonlinear term $\chi|\Psi|^4$ considered here represents an effective three-body, or quintic, interaction that becomes relevant beyond the dilute-polariton regime. Microscopically, such nonlinearities can arise from phase-space filling, excitonic saturation, polarization mixing, and higher-order polariton scattering processes at elevated densities~\cite{Carusotto2013, SedovE}. These mechanisms lead naturally to an effective cubic-quintic description of the polariton condensate, in which the quintic contribution captures the leading correction to the conventional binary interaction~\cite{SedovE}. A phenomenological description of saturation can be written as
\begin{align}
V_{\mathrm{eff}}(n)\simeq g_C n\left(1-\frac{n}{n_{\mathrm{sat}}}\right),
\end{align}
where $n_{\mathrm{sat}}$ denotes an effective saturation density. Comparing this expression with the nonlinear potential $g_C|\Psi|^2+\chi|\Psi|^4$ gives the estimate
\begin{align}
\chi\simeq-\frac{g_C}{n_{\mathrm{sat}}},
\end{align}
for saturation-induced corrections. Thus, the magnitude of the effective higher-order coupling is expected to scale as $|\chi|\sim g_C/n_{\mathrm{sat}}$, with its precise value depending on the microscopic origin of the nonlinear correction. In terms of the dimensionless background density $n_0$, we define the dimensionless three-body interaction strength as
\begin{align}
\bar{\chi}=\frac{\chi n_0}{g_C}.
\end{align}
For representative condensate densities, $n_0\sim10$--$50~\mu\mathrm{m}^{-2}$, this scaling motivates values of $|\bar{\chi}|$ in the range $10^{-2}$--$10^{-1}$ for weak higher-order nonlinearities.

Our aim, however, is not to model a particular experimental device, but rather to characterize the generic effects of effective three-body interactions on the nonlinear dynamics and stability of topological defects in driven-dissipative condensates. The parameters varied in our simulations are dimensionless combinations of the underlying physical scales and therefore do not, in general, correspond to independent experimental control parameters. Their variation can instead represent changes in several experimentally relevant quantities, including the cavity quality factor, exciton-photon detuning, reservoir relaxation rate, pump strength, and material composition. In this sense, the dimensionless parameter $\bar{\chi}$ provides a convenient measure of the relative importance of higher-order interactions compared with the conventional binary polariton interaction. Values $\bar{\chi}=\pm0.1$ correspond to a weak higher-order nonlinear correction and are within the regime motivated by experimentally relevant saturation scales. We also consider larger values, $\left|\bar{\chi}\right|\sim0.5$--$1$, in order to explore the broader theoretical regime in which three-body interactions substantially modify dark-soliton stability and the transverse snake instability. Such values should therefore be interpreted as a generalized parameter regime rather than as a direct representation of a specific existing microcavity.

The coupled equations (\ref{eq:Psi2}) and (\ref{eq:rate}) describe the driven-dissipative dynamics of the condensate wavefunction and the reservoir density in the presence of nonlinear interactions, gain, and loss. To gain insight into the fundamental behavior of the system, it is useful to first analyze the stationary solutions that arise from the balance between pumping and dissipation. In particular, we consider the case of spatially uniform and time-independent pumping, for which the system can support a homogeneous nonequilibrium steady state. The properties of this steady state form the basis for understanding the subsequent dynamical behavior of the condensate.

For notational brevity, the overbars on all dimensionless fields ($\Psi, n_R, \psi, m_R$) and dimensionless parameters ($\chi, g_R, \gamma_C, \gamma_R, R, P_0, P_{\mathrm{th}}$) are omitted in the subsequent sections.

\subsection{Steady-state under uniform continuous pumping}
\label{sec:steady}
The driven-dissipative nature of the condensate allows the system to reach a nonequilibrium steady state determined by the balance between gain and loss. To characterize this state, we consider the case of spatially uniform and continuous pumping and determine the corresponding homogeneous stationary solutions. Under spatially uniform and time-independent pumping $P(\vec{r},t)=P_0$, the system supports homogeneous stationary solutions of the form
\begin{align}
\Psi(\vec{r},t) = \sqrt{n_{C}^{0}} \, e^{-i\mu t}, \;\; n_R(\vec{r},t) = n_R^0,
\label{psi-homogeneous}
\end{align}
where  $|\Psi_{0}|^{2}=n_{C}^{0}$ is real, $n_R^0$ is constant, and $\mu$ is the dimensionless chemical potential.

Substituting \eqref{psi-homogeneous} into the dimensionless equations \eqref{eq:model} yields the following algebraic conditions
\begin{subequations}
\begin{align}
\mu \Psi_0 &= \bigl( \Psi_0^2 + g_R n_R^0 \bigr) \Psi_0 + \frac{i}{2} (R n_R^0 - \gamma_C) \Psi_0, \\
0 &= P_0 - (\gamma_R + R \Psi_0^2) n_R^0.
\end{align}
\end{subequations}
For a non-vanishing condensate amplitude $\Psi_0 > 0$, the imaginary part of the first equation requires the gain-loss balance
\begin{align}
%R n_R^0 = \gamma_C \quad \Rightarrow \quad 
n_R^0 = \frac{\gamma_C}{R}.
\end{align}
This clamping of the reservoir density above threshold is a hallmark of polariton condensation~\cite{Deng2010, Carusotto2013, Kasprzak2006}.

Inserting the clamped value into the second (reservoir) equation determines the condensate density:
\begin{align}
\Psi_0^2 = \frac{P_0 - P_{\rm th}}{\gamma_C}, \qquad P_{\rm th} \equiv \frac{\gamma_C \gamma_R}{R},
\end{align}
where $P_{\rm th}$ is the dimensionless threshold pump intensity. The real part of the wave-function equation then gives the chemical potential
\begin{align}
\mu = \Psi_0^2 + g_R \frac{\gamma_C}{R}.
\end{align}
Below threshold ($P_0 < P_{\rm th}$), the only physical solution is the empty state $\Psi_0 = 0$, in which case the reservoir density reduces to the linearly pumped value $n_R^0 = P_0 / \gamma_R$ and the chemical potential (formally obtained by taking the zero-density limit) is $\mu = g_R P_0 / \gamma_R$.

The complete steady-state solution above threshold therefore reads
\begin{align}
 \Psi_0 = \sqrt{\frac{P_0 - P_{\text{th}}}{\gamma_C}}, \quad
 n_R^0 = \frac{\gamma_C}{R}, \quad
 \mu = \frac{P_0 - P_{\text{th}}}{\gamma_C} + \frac{g_R \gamma_C}{R}.
\end{align}
Above threshold, the reservoir density is pinned at a constant value independent of pumping strength, while the condensate density increases linearly with the excess pump intensity \(P_0 - P_{\text{th}}\). 
This nonlinear threshold behavior and reservoir clamping are routinely observed in experiments~\cite{Kasprzak2006, Bajoni2008} and constitute the defining signature of Bose-Einstein condensation in driven-dissipative polariton systems.

\subsection{Hydrodynamic Interpretation of the Snake Instability}

An intuitive understanding of the snake instability can be gained within the hydrodynamic representation of the driven-dissipative Gross-Pitaevskii equation. Using the Madelung transformation
\begin{align}
\Psi(\vec{r},t) = \sqrt{n_{C}^{0}}\, e^{i S(\vec{r},t)},
\end{align}
the superfluid density $n_{C}^{0}=n(x,y)=|\Psi|^2$ and velocity field $\vec v(\vec r,t) = \frac{\hbar}{M}\nabla S(\vec r,t)$ obey continuity and Euler-type equations modified by gain, loss, and reservoir-mediated interactions~\cite{Carusotto2013, Wouters2007, Fetter2001}. The current density of the wavefunction can be written as
\begin{align}
    \vec{j}(\vec{r},t) = \Big(\frac{\hbar}{2Mi}\Big)[\Psi^{*}\nabla \Psi- (\nabla \Psi^{*})\Psi].
    \label{eq:current-density}
\end{align}
The current density takes the hydrodynamic form as 
\begin{align}
    \vec{j}(\vec{r},t) = n(\vec{r},t) \, v(\vec{r},t).
    \label{eq:hydro-J}
\end{align}
Here, the rotational velocity $v(\vec{r}, t)$ can be expressed in terms of velocity potential $\Phi(\vec{r},t)$
\begin{align}
    \vec{v}(\vec{r},t) = \nabla \Phi(\vec{r},t),
    \label{eq:irr-velocity}
\end{align}
where the velocity potential is 
\begin{align}
    \Phi(\vec{r},t) = \frac{\hbar}{M} S(\vec{r},t).
    \label{eq:Phi}
\end{align}
Taking the gradient on both sides of Eq.~\eqref{eq:Phi}, we have
\begin{align}
    \nabla \Phi(\vec{r},t) = \frac{\hbar}{M}\nabla S(\vec{r},t),
\end{align}
which implies
\begin{align}
    \bm v(\vec r,t) = \frac{\hbar}{M}\nabla S(\vec{r},t).
    \label{eq:velocity}
\end{align}
Using the Madelung transformation of the condensate wavefunction, the velocity field can equivalently be written as
\begin{align}
    \bm v(\vec r,t)
    = \frac{\hbar}{M}\nabla S(\vec{r},t)
    = \frac{\hbar}{M}\frac{\mathrm{Im}\!\left(\Psi^{*}\nabla\Psi\right)}{|\Psi|^{2}}.
\end{align}
Assuming a uniform background density $n(x,y) = n_0$, the wavefunction is given by
\begin{align}
    \Psi(x,y,t) = \sqrt{n_0} \left[ \tanh\left(\frac{x}{\xi_s}\right) + \mathrm i A(t) \frac{\sin(ky)}{\cosh(x/\xi_s)} \right], \label{eq:trial_wave_corrected}
\end{align}
The velocity components $v_{x}$ and $v_{y}$ [See Appendix~\ref{appendix:velocity_derivation} for details] are given by %
\begin{subequations}
\label{del-s}
\begin{align}
v_x & = -\frac{A(t)}{\xi}\left[\sech^{3}\!\left(\frac{x}{\xi}\right) + \tanh^{2}\!\left(\frac{x}{\xi}\right)\sech\!\left(\frac{x}{\xi}\right)\right]\sin(ky), \\
v_y &= k A(t)\,\tanh\!\left(\frac{x}{\xi}\right)\sech\!\left(\frac{x}{\xi}\right)\cos(ky).
\end{align}
\end{subequations}%
These expressions show that the transverse modulation of the soliton introduces spatially varying velocity components along both the longitudinal ($x$) and transverse ($y$) directions. In particular, the dependence of $\sin(ky)$ and $\cos(ky)$ indicate that the velocity field develops periodic variations along the soliton line. As the instability amplitude $A(t)$ grows, these transverse velocity perturbations become increasingly pronounced, leading to bending and deformation of the initially straight soliton.

\begin{figure}[!ht]
\centering\includegraphics[width=\linewidth]{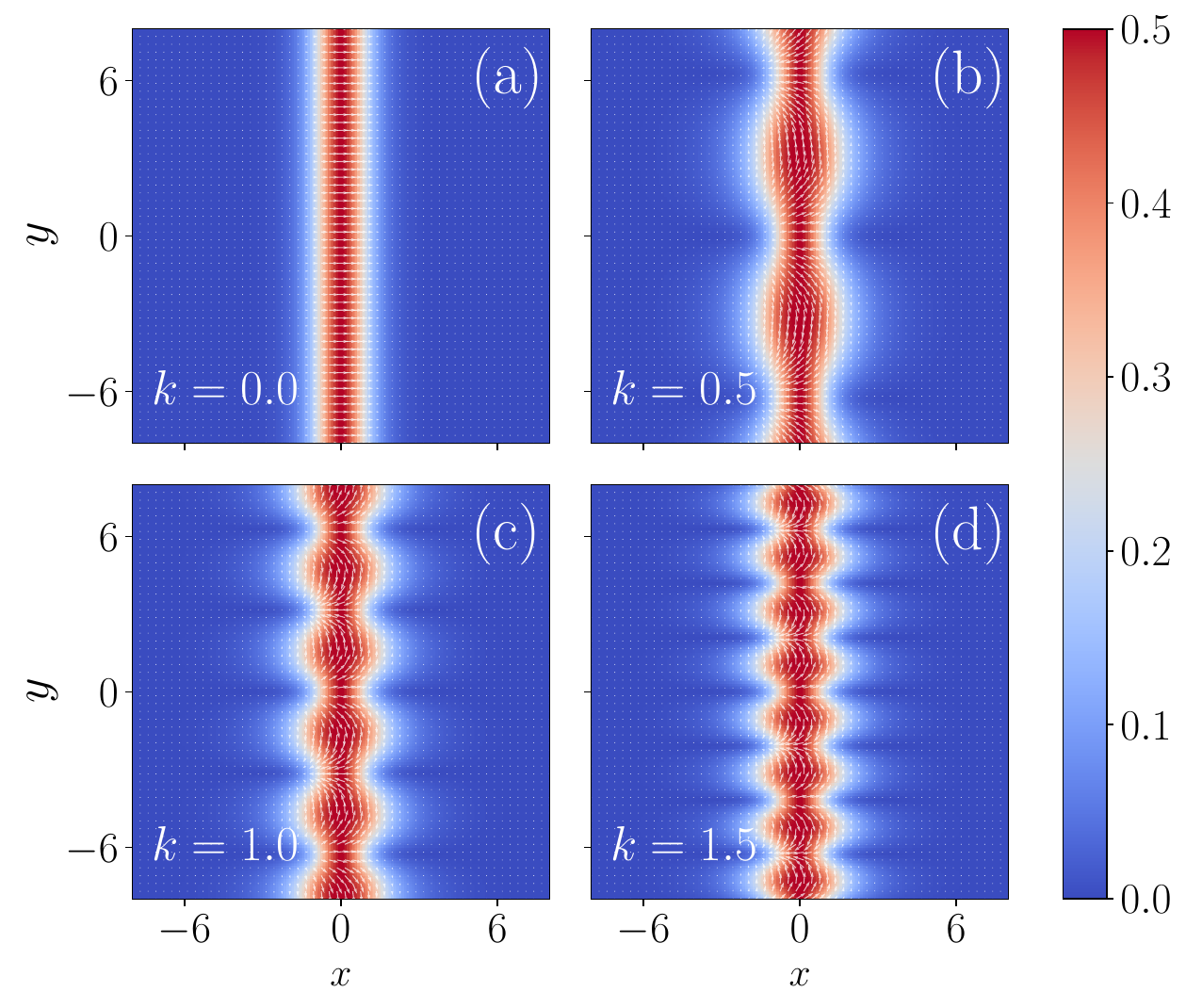}
\caption{The velocity fields from Eq.~\eqref{del-s} for an instability amplitude $A(0) = 0.25$ for different $k$. }
\label{fig:dels}
\end{figure}

Figure~\ref{fig:dels} illustrates the velocity fields for different wave numbers ($k$) at a fixed instability amplitude $A=0.25$. %
With increasing $k$, the spatial modulation of the velocity field becomes stronger, resulting in enhanced transverse flows around the soliton core. 

From a hydrodynamic perspective, these growing transverse flows signal the breakdown of the planar soliton structure and the onset of the snake instability. Consequently, the soliton becomes dynamically unstable and eventually decays into pairs of vortices, which correspond to localized phase singularities in the condensate. As the instability develops further, the soliton stripe breaks up and nucleates vortex-antivortex pairs, which appear as localized phase singularities in the condensate and represent the final stage of the snake instability.

\section{Asymptotic description of the dynamics of the two-dimensional dark solitons}
\label{sec:AsympDark}

Several experimental and theoretical methods exist for generating and studying vortices in BECs \cite{Matthews1999, Fetter2001, Aftalion2006, Flayac2010, White2014, Gnusov2023, Gnusov2024}. These include phase imprinting, rotating traps, stirring with laser beams, and dynamical instabilities of nonlinear excitations. Among these mechanisms, the decay of a dark soliton via the transverse (snake) instability provides a particularly direct route for vortex nucleation in effectively two-dimensional geometries. This process is not only experimentally accessible but also theoretically appealing, as it connects soliton physics to vortex dynamics through a well-defined instability mechanism. In one spatial dimension, dark solitons are robust nonlinear excitations characterized by a localized density depletion and phase jump across the nodal plane. However, in two dimensions, such planar structures are dynamically unstable against transverse perturbations. Small modulations along the direction perpendicular to propagation grow exponentially, leading to the breakup of the soliton stripe into vortex-antivortex pairs. This instability mechanism, commonly referred to as the snake instability, plays a central role in the formation of coherent vortex structures from initially non-topological excitations.

To capture the early stages of this dynamics and the subsequent evolution of the soliton filament, it is useful to develop an asymptotic description valid in the regime where the soliton width is small compared to the characteristic length scale of the background inhomogeneity~\cite{Flayac2010, Jia2025}. In this framework, the dark soliton is treated as a quasi-one-dimensional object embedded in a slowly varying two-dimensional condensate background. The separation of scales between the soliton core (of order of the healing length) and the external trapping potential allows for a systematic reduction of the Gross-Pitaevskii equation to an effective equation governing the soliton centerline dynamics.

In the asymptotic description, the wavefunction of a two-dimensional dark soliton in an inhomogeneous system can be written as
\begin{align}
 \psi(\vec{r},t) = G(\vec{r}) F(\vec{r},t),
 \label{eq:psi_asym}
\end{align}
where $G(\vec{r})$ represents the solution of the unperturbed condensate under the influence of the external potential $U(\vec{r})$, while $F(\vec{r},t)$ describes the time evolution of finite-amplitude perturbations on an initially inhomogeneous condensate. In the absence of the reservoir density perturbation, Eq.~\eqref{eq:Psi2} reduces to a nonlinear Schr\"{o}dinger equation of the following form 
\begin{align}
    \mathrm{i} \frac{\partial \psi}{\partial t} &+ \frac{1}{2}{\nabla^{2}} \psi+ (1-|\psi|^{2})\psi+ \chi (1-|\psi|^{4})\psi =0 
    \label{eq:psi-nlse}
\end{align}
which admits  dark soliton solution in the form of Eq.~\eqref{eq:trial_wave_corrected}. Substituting the ansatz \eqref{eq:psi_asym} into \eqref{eq:psi-nlse}, we obtain a set of coupled equations for the unperturbed and perturbed components~\cite{smirnov2012}, 
\begin{align} 
 \frac{1}{2} \nabla^{2} G + (1-|G|^{2})G + \chi(1-|G|^{4})G = 0 
\end{align}
\begin{align}
 i \frac{\partial F}{\partial t} &+\frac{{\nabla^{2}} F}{2} + |G|^{2}\left(1-|F|^{2}\right)F \notag \\ & 
 +\chi|G|^{4}\left(1-|F|^{4}\right)F=-\frac{1}{G} \nabla G \cdot \nabla F
\end{align}
Assuming that the unperturbed component $G(\vec{r})$ and the external potential $U(\vec{r},t)$ vary slowly within the localization region, the dynamics of the two-dimensional quasi-soliton in a condensate at rest can be described by
\begin{align}
 G(\vec{r}) = g(\vec{r}) \exp(i\theta)
\end{align}
where $\theta$ is a constant condensate phase and $g(\textbf{r})$ is a real-valued function of the position. The position vector is $\vec{r} = x \hat{x}+ y \hat{y}$, with $\hat{x}$ and $\hat{y}$ denoting the cartesian unit vectors. To describe the soliton geometry more conveniently, we transform from Cartesian coordinates to orthogonal curvilinear coordinates (s,$\eta$)  as 
\begin{align}
\vec{r} = \vec{r}_{0}(s) + \eta \vec{n_{0}}(s)
\end{align}
where, $\vec{r}_0 (s)$ specifies the centre of the dark 
soliton  and $\vec{n_0}(s)$ is the local normal vector. In these coordinates, the condensate equation can be reformulated in a natural way for analyzing the soliton dynamics. Now, the condensate equation in the orthogonal curvilinear coordinates employing Lame coefficients can be written as
\begin{align}
	\mathrm{i} \frac{\partial F}{\partial t} + & \frac{1}{2h_{s}h_{\eta}}\left[\frac{\partial}{\partial s}\left(\frac{h_{\eta}}{h_{s}}\frac{\partial F}{\partial s}\right)+\frac{\partial}{\partial \eta}\left(\frac{h_{s}}{h_{\eta}}\frac{\partial F}{\partial \eta}\right)\right] \notag \\  & 
    +|G|^{2}\left(1-|F|^{2}\right)F + \chi |G|^{4}\left(1-|F|^{4}\right)F \notag \\ & \quad = \frac{1}{G}\left[\frac{1}{h_{s}^{2}} \frac{\partial G}{\partial s}\frac{\partial F}{\partial s} +\frac{1}{h_{\eta}^{2}} \frac{\partial G}{\partial \eta}\frac{\partial F}{\partial \eta}\right]
\end{align}
We expand the undisturbed function $|G(\mathbf{r})|^{2}$ near the curve $r_{0}(s)$ in the form of a Taylor series in terms of $\eta$ as 
\begin{align}
	|G(\vec{r})|^{2} &= |G(\vec{r}_{0}+\eta \vec{n}_{0})|^{2} 
\end{align}
Upon expanding $|G(\vec{r})|$ at $\eta=0$, we obtain 
\begin{align}
    |G(\vec{r})|^{2}= |G_{0}|^{2} + \eta \frac{\partial |G|^{2}}{\partial \eta}\Bigg|_{\eta=0} + \frac{\eta^{2}}{2} \frac{\partial^{2} |G|^2}{\partial \eta^{2}}\Bigg|_{\eta=0}
\end{align}
The transformed equation can be written as 
\begin{align}
    \mathrm{i} \frac{\partial F}{\partial t}  &+\frac{1}{2}\frac{\partial^{2} F}{\partial s^{2}}+ \frac{1}{2}\frac{\partial^{2} F}{\partial \eta^{2}} +|G(r_{0})|^{2}(1-|F|^{2})F \notag \\ &
    +\chi|G(r_{0})|^{4}(1-|F|^{4})F  %\notag \\ & 
    =-\frac{\kappa \eta}{2} \frac{\partial^{2} F}{\partial s^{2}} %\notag\\& 
    +\frac{\kappa}{2}\frac{\partial^{2} F}{\partial \eta^{2}} \notag \\ & -\Bigg[\frac{\partial |G|^{2}}{\partial \eta }\Bigg]_{\eta =0} \eta (1-|F|^{2})F %\notag \\ &  
    - \chi\Bigg[\frac{\partial |G|^{4}}{\partial \eta }\Bigg]_{\eta =0} \eta (1-|F|^{4})F \notag \\ &
    -\Bigg[\frac{1}{G} \frac{\partial G}{\partial s}\Bigg]\Bigg|_{\eta=0}\frac{\partial F}{\partial s} %\notag \\&
    -\Bigg[\frac{1}{G} \frac{\partial G}{\partial \eta}\Bigg]\Bigg|_{\eta=0}\frac{\partial F}{\partial \eta} + \mathcal{O}(\lambda^{2}) 
    \label{eq:inhomo-taylor}
\end{align}
After this transformation, the solution of  equation \eqref{eq:inhomo-taylor} near the curve $r_{0}(s)$ can be represented as an asymptotic series of $\lambda$,
with the instability amplitude  slowly varying in time given by 
$A(t) = A(\lambda t)$ where $\lambda\ll 1$. Under these circumstances, the solutions can be expressed in the form of asymptotic expansion in $\lambda$ with $\xi = x/ \xi_{s}$ as
\begin{subequations}
    \begin{align}
	\psi(\vec{r},t)= \psi_{s}(\xi,y, A(\lambda t))+ \sum_{j=1}^{\infty} \lambda^{j}\psi_{j}(\xi,y, A(\lambda t)) \label{order-psi}\\
	m_{R}(\vec{r},t)= \lambda m_{R}^{0}(\xi,y, \lambda t)+ \sum_{j=1}^{\infty} \lambda^{j+1} m_{Rj}(\xi,y,\lambda t) \label{order-mr}
\end{align}
\end{subequations}
The small parameter $\lambda$ is introduced to organize the asymptotic perturbation expansion enabling us quantify  the separation of scales between the intrinsic soliton dynamics and the slow reservoir response. In the general formulation, $\lambda$ should be regarded as an abstract ordering parameter rather than a fixed combination of physical parameters.

We now substitute Eq.~\eqref{order-psi} into \eqref{eq:Psi2} and consider only the terms  upto the first order in $\lambda$ with  the linear inhomogeneous equation $\psi_{1}(\xi, \,y,\lambda t)$ taking the following form 
\begin{align}
    \frac{1}{2}\frac{\partial^{2}}{\partial \xi^{2}} & + \frac{1}{2}\frac{\partial^{2}}{\partial y^{2}}- \left[\left(1-2|\psi_{s}|^{2}\right)+ \chi \left(1-3|\psi_{s}|^{4}\right)\right] \notag \\ &
    + [1+2 \chi|\psi_{s}|^{2}]\psi_{s}^{2}\psi_{1}^{*} = \mathrm{i}\frac{\partial A}{\partial t}+ \frac{\partial \psi_{s}}{\partial A}- \mathcal{R}(m_{R}^{0}, \psi_{s})
\end{align}
From this, the Fredholm alternative can be defined as 
\begin{align}
	 Re\Bigg[\int_{-\infty}^{\infty}\Bigg(\mathrm{i}\frac{\partial A}{\partial t}+ \frac{\partial \psi_{s}}{\partial A}- \mathcal{R}(m_{R}^{0}, \psi_{s})\frac{\partial \psi_{s}^{*}}{\partial \xi} \Bigg)d\xi \Bigg]=0.
	\label{R_tot}
\end{align}
Equation \eqref{R_tot} leads to the dynamical equation of the following form 
\begin{align}
    \frac{d\mathcal{E}_{s}}{dt} = \int_{-\infty}^{+\infty} d \xi \, dy \left(\mathcal{R}(m_{R}^{0}, \psi_{s})\frac{\partial \psi_{s}^{*}}{\partial \xi}+ \mathcal{R}^{*}(m_{R}^{0}, \psi_{s})\frac{\partial \psi_{s}}{\partial \xi}\right).
    \label{eq:dyn-ode}  
\end{align}
The energy of the dark soliton will now become
\begin{align}
\mathcal{E}_s
=\int d \xi\,dy\left[\frac{1}{2}|\nabla \psi|^{2}+\frac{1}{2}\left(1-|\psi|^{2}\right)^{2}+\frac{\chi}{3}\left(1-|\psi|^{2}\right)^{3}\right]
\label{eq:energy}
\end{align}
\begin{align}
    \mathcal{E}_s = A^2 \left(\frac{k^2 \xi }{2}-\frac{8 \xi  \chi }{15}-\frac{2 \xi }{3}+\frac{1}{6 \xi }\right)-\frac{1}{9} \xi  \chi  A^6 \notag \\
    +A(t)^4 \left(\frac{2 \xi  \chi }{5}+\frac{\xi }{4}\right)+\frac{16 \xi  \chi }{45}+\frac{2 \xi }{3}+\frac{2}{3 \xi }
\end{align}
Since the instability amplitude $A(t)$ is very small,  we can neglect the higher order powers of instability amplitude  to obtain stable vortex solutions as   
\begin{align}
    \mathcal{E}_s = A^2 \left(\frac{k^2 \xi }{2}-\frac{8 \xi  \chi }{15}-\frac{2 \xi }{3}\right)+\frac{16 \xi  \chi }{45}+\frac{2 \xi }{3}+\frac{2}{3 \xi }
\end{align}

\subsection{Weak-pumping region} 
%%%%%%%%%%%%%%%%%%%%%%%%%%%%%%%%%%%%%%%%%%%%%%%%%%%%%%%
% Weak pumping 
%%%%%%%%%%%%%%%%%%%%%%%%%%%%%%%%%%%%%%%%%%%%%%%%%%%%%%%
In order to understand the influence of the reservoir on the transverse instability of the dark soliton, we now analyze the dynamics in the weak-pumping regime. Physically, this regime corresponds to pumping strengths slightly above the condensation threshold, where the reservoir density remains small, and its response to condensate fluctuations can be treated perturbatively. This condition can be expressed as
\begin{align}
    \frac{P_{0}}{P_{\mathrm{th}}}-1 \ll 1.
    \label{eq:weak_pump}
\end{align}
In this limit, the relaxation rate of the reservoir is typically much larger than the characteristic timescale of the condensate dynamics. As a result, the reservoir adjusts rapidly to variations in the condensate density and can be treated within an adiabatic approximation. Consequently, the spatial scale of the reservoir perturbation $m_{R}^{0}(\xi,y,\lambda t)$ becomes comparable to the width of the dark soliton, $\xi_{s}$. 
From the reservoir rate Eq.~\eqref{eq:rate} together with the asymptotic expansion \eqref{order-mr}, the leading-order reservoir perturbation is directly coupled to the condensate density $|\psi_{s}(\xi,y,\lambda t)|^{2}$. In the weak-pumping limit, the ratio of the condensate decay rate to the reservoir relaxation rate is small, leading to 
\begin{align*}
    \lambda = \frac{\gamma_{C}}{\gamma_{R}} \ll 1.
\end{align*}
The  leading order contribution to  the reservoir density perturbation is therefore given by 
\begin{align}
    m_{R}^{0}
    = \frac{\gamma_{C}}{\gamma_{R}}\left(1-|\psi_{s}|^{2}\right).
    \label{eq:mR_leading}
\end{align}
where the dark soliton ansatz is defined in transformed coordinates with background density $n$ as 
\begin{align}
   \psi_{s}(\xi,y,t) = \sqrt{n} \left[ \tanh\left(\xi\right) + \mathrm i A(t) \frac{\sin(ky)}{\cosh(\xi)} \right]
\end{align}
Substituting Eq.~\eqref{eq:mR_leading} into the dynamical Eq.~\eqref{eq:dyn-ode} and using the dark-soliton ansatz $\psi_{s}(\xi,y,t)$  defined above, we obtain the evolution equation for the instability amplitude $A(t)$,
\begin{align}
    \dot{A} = \alpha A^2, 
    \label{ode-weak}
\end{align}
where
\begin{align*}
    \alpha = \frac{\sqrt{n}\pi^{2} R \xi_{s} \gamma_{C}}{\gamma_{R}}\left(1-\frac{n}{2}\right).
\end{align*}
The solution of the above ordinary differential equation \eqref{ode-weak} can be written as 
\begin{align}
A(t) = \frac{A_{0}}{1-\alpha A_{0}t}
\end{align}
To find the disappearance/lifetime of the stable vortices, we need to have 
\begin{align}
1-\alpha A_{0}t =0
\end{align}
From this, we can obtain the lifetime $\tau$ of the vortices as:
\begin{align}
\tau = \frac{1}{\alpha A_{0}}
\end{align}
From this equation, the disappearance of the stable vortices is determined by 
\begin{align}
    \frac{1}{\tau} = \frac{1}{2}\left(n-\frac{n^{3/2}}{2}\right)\left(\frac{A_{0}\pi^{2} R\xi_{s}\gamma_{C}}{\gamma_{R}}\right)
    \label{life-time-weak}
\end{align}
The disappearance of the stable vortex is determined by the lifetime ($\tau$) of the dark soliton as 
\begin{align}
    \tau = \frac{2 \tau_{0}}{A_{0}\pi^{2}(n-\frac{n^{3/2}}{2})\gamma_{C}}\left(\frac{P_{\text{th}}}{P_{0}-P_{\text{th}}}\right)
    \label{lifetime-tau}
\end{align}
The above analysis shows that in the weak-pumping regime, the reservoir responds adiabatically to the condensate density, and the transverse instability of the dark soliton is governed by the nonlinear evolution equation for the instability amplitude $A(t)$. The resulting growth of the perturbation determines the lifetime of the vortices formed through the snake instability, which depends explicitly on the ratio $\gamma_C/\gamma_R$ and the pumping strength relative to the threshold. In this regime, the dynamics are dominated by the fast relaxation of the reservoir, leading to a relatively simple relation between the instability growth and the condensate parameters.

However, as the pumping strength increases further above the threshold, the assumption of fast reservoir relaxation is no longer valid. In this regime, the reservoir dynamics become comparable to the condensate timescale and begin to significantly influence the evolution of the instability. It is therefore necessary to analyze the system in the strong-pumping regime, where the reservoir response and the resulting soliton dynamics exhibit qualitatively different behavior. This regime is considered in the following section.

\subsection{Strong-pumping regime}
\label{sec:strong}
In contrast to the weak-pumping regime discussed above, we now consider the opposite limit of strong pumping characterized by
\begin{align}
    \frac{P_{0}}{P_{\mathrm{th}}} \gg 1.
    \label{con-strong}
\end{align}
In this regime, the reservoir density is no longer slaved perturbatively to the condensate density, and its dynamics must be treated more explicitly. Retaining terms up to first order in the asymptotic expansion parameter $\lambda$, the reservoir rate Eq.~\eqref{eq:rate} is conveniently expressed
in the moving coordinate
\begin{align}
\zeta = x - \phi(y,t),
\label{zeta-transform}
\end{align}
which follows the transverse deformation of the soliton. Under this coordinate transformation, the time derivative becomes
\begin{align}
\partial_t m_R^0 = -\phi_t\,\partial_\zeta m_R^0.
\label{timederiv-mr}
\end{align}
Substituting Eq.~\eqref{zeta-transform} into the reservoir equation yields
\begin{align}
    {-\phi_t} \frac{\partial m_R^0}{\partial \zeta}
    + \bigl(\gamma_R + R|\psi_s|^2\bigr)m_R^0
    = \gamma_C\bigl(1-|\psi_s|^2\bigr).
\end{align}
This provides a first-order ordinary differential equation governing the leading-order reservoir perturbation,
\begin{align}
    \frac{\partial m_R^0}{\partial \zeta}
    + \frac{\gamma_R + R|\psi_s|^2}{\eta_t}\, m_R^0
    = \frac{\gamma_C\bigl(1-|\psi_s|^2\bigr)}{\eta_t}.
    \label{eq:ode-m0r}
\end{align}
Thus, in the strong-pumping regime, the reservoir polariton density responds locally to the condensate wavefunction, but its spatial variation is also modulated by the soliton deformation through the moving coordinate $\zeta$.

The solution of the reservoir [Eq.~\eqref{eq:ode-m0r}] can be written as
\begin{align}
    m_R^0(\xi,y,t) = & -\frac{\gamma_C}{R} + \Delta m \exp\Bigg[ -\frac{R n}{\eta_t} \Big( \xi - \tanh(\xi) \notag \\ & + A^2 \sin^2(ky) \tanh(\xi) \Big) \Bigg]
    \label{eq:reservoir-solution}
\end{align}
Substituting  $m_R^0$ in \eqref{R_tot}, we obtain the dynamical equation of $A(t)$  
\begin{align}
    \frac{dA}{dt} = \left( \frac{R}{4} \mathcal{J} - \frac{\gamma_C}{2} \right) A
    \label{eq:amplitude-raw}
\end{align}
where $\mathcal{J}$ is the nonlocal integral.  Solving the nonlocal term will result in a rate of change of instability amplitude as
(For $\lambda \gg 1$, using $\mathcal{K}(\lambda) \sim \sqrt{\pi/\lambda} = \sqrt{\pi\eta_t/(Rn)}$:)
\begin{align}
    \frac{dA}{dt} = \gamma_C \left[ -1 + \frac{1}{4} \sqrt{\frac{\pi\eta_t}{Rn}} \right] A.
    \label{eq:amplitude-asymptotic}
\end{align}
Using  Eq.~\eqref{eq:amplitude-asymptotic}, the life time of the condensate can be expressed as 
\begin{align}
    \tau = \frac{1}{\left({\gamma_C \left| -1 + \frac{1}{4} \sqrt{\frac{\pi \eta_t}{R n}} \right|}\right)}.
    \label{eq:tau-final}
\end{align}

\section{Numerical results}
\label{sec:num}
Having examined the dynamics of exciton-polariton condensates in both weak- and strong-pumping regimes using the asymptotic description in the previous section, we now investigate their behavior numerically by solving the dissipative GP equation \eqref{eq:Psi2} coupled with the reservoir rate equation \eqref{eq:rate}. The condensate dynamics are computed using the split-step Crank-Nicolson method~\cite{Muruganandam2009, Vudragovic2012, KishorKumar2015, Ravisankar2021}, while the reservoir evolution is integrated using a fourth-order Runge-Kutta method. In the simulations, the spatial discretization is chosen as \(dx = dy = 0.05\), and the time step is \(dt = 10^{-4}\pi\).

To investigate the transverse snake instability, the initial state at \(t = 0\) is configured as a dark soliton perturbed transversely along the \(y\)-direction with wavenumber \(k = 1.0\) and chemical potential \(\mu = 1.0\).  %
\begin{figure}[!ht]
\centering
\includegraphics[width=\linewidth]{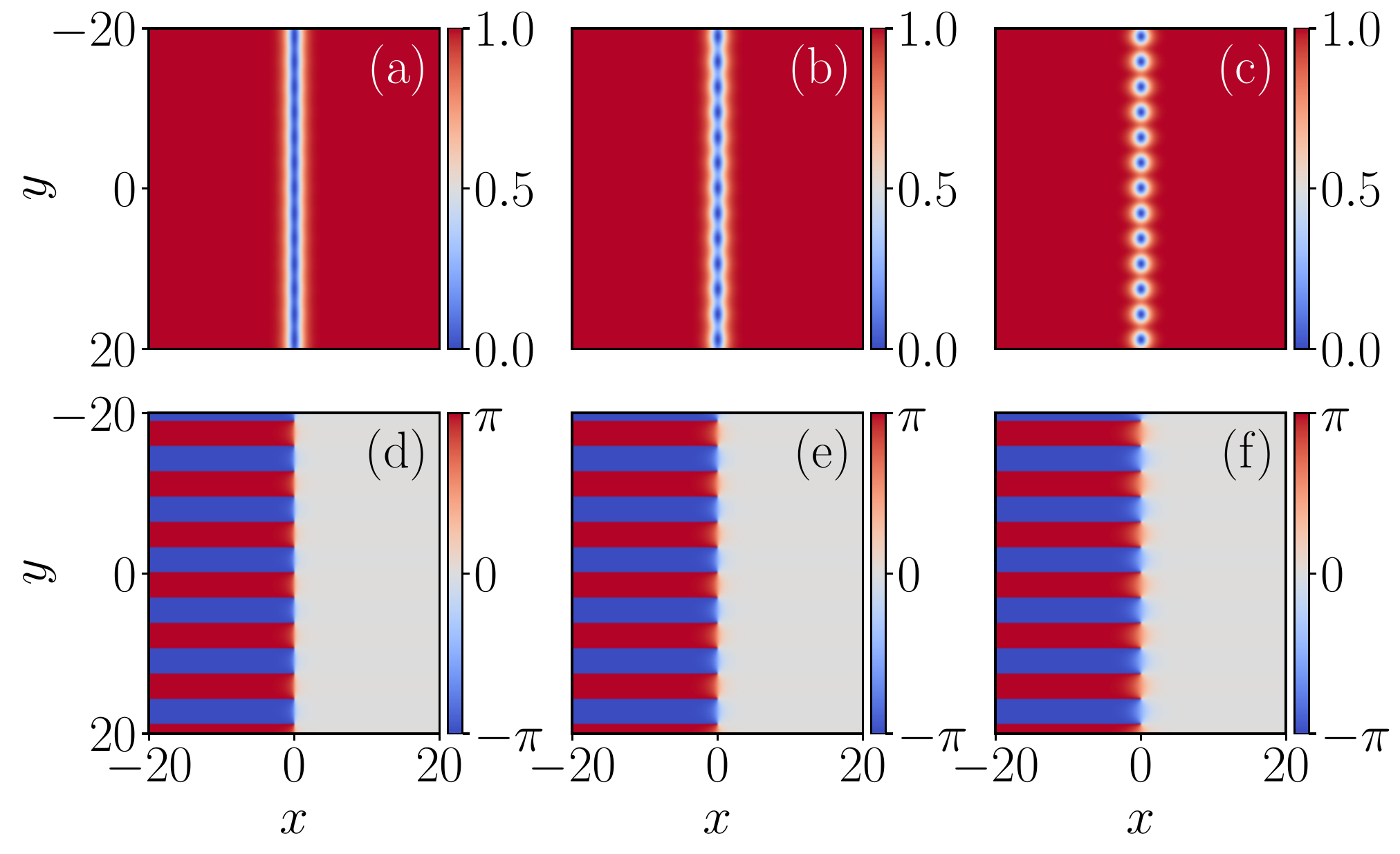}
\caption{Initial density and phase profiles of a modulated dark soliton at \(t = 0\) for wavenumber \(k = 1.0\) and chemical potential \(\mu = 1.0\). Top row: density profiles for perturbation amplitudes (a) \(A = 0.25\), (b) \(A = 0.5\), and (c) \(A = 0.9\). Bottom row: The associated phase profiles shown in (d)--(f) for the same respective amplitudes.}
\label{fig:initial_profiles}
\end{figure}
The initial density and phase distributions for various perturbation amplitudes (\(A = 0.25, 0.5, 0.9\)) are shown in Fig.~\ref{fig:initial_profiles}. The dynamical simulations utilize all these cases (\(A = 0.25, 0.5, 0.9\)) in this study to thoroughly examine the evolution across different perturbation strengths. As \(A\) increases, the density minimum along the soliton stripe becomes shallower and transverse modulations become more pronounced [cf. Figs.~\ref{fig:initial_profiles}(a)--(c)]. Simultaneously, the phase profiles display sharper localized gradients across the nodal line [cf. Figs.~\ref{fig:initial_profiles}(d)--(f)], which prepares the system for the subsequent nonlinear breakup of the dark soliton into vortex pairs.

%\subsection{Instability in the absence of pumping}
\subsection{Instability dynamics in the absence of pumping}

Before analyzing the influence of pumping and reservoir dynamics on the transverse instability, it is useful to first examine the behaviour of the system in the absence of the pumping. In this limit, the condensate evolves without gain or loss mechanisms and the reservoir density is set to zero ($m_R=0$). Studying this limit allows us to isolate the intrinsic instability mechanism of a two-dimensional dark soliton without the influence of external driving or reservoir-mediated effects.

In a conservative condensate, a planar dark soliton is dynamically unstable against transverse perturbations. Small modulations along the direction parallel to the soliton stripe grow with time through the well-known snake instability. As the instability develops, the initially straight soliton begins to bend and eventually breaks into a sequence of vortex-antivortex pairs. This process represents a fundamental pathway for the generation of quantized vortices from non-topological excitations and plays a central role in the nonlinear dynamics of two-dimensional quantum fluids. Investigating this regime therefore provides an important reference point for understanding the role of higher-order interactions and nonlinear effects in the evolution of the instability. In particular, it allows us to identify the baseline dynamics of the soliton decay and the formation of vortex structures before additional mechanisms such as pumping and reservoir coupling are introduced.
 
\begin{figure}[!ht] 
%\centering\includegraphics[width=\linewidth]{dyn-fin/fig02}
\centering\includegraphics[width=\linewidth]{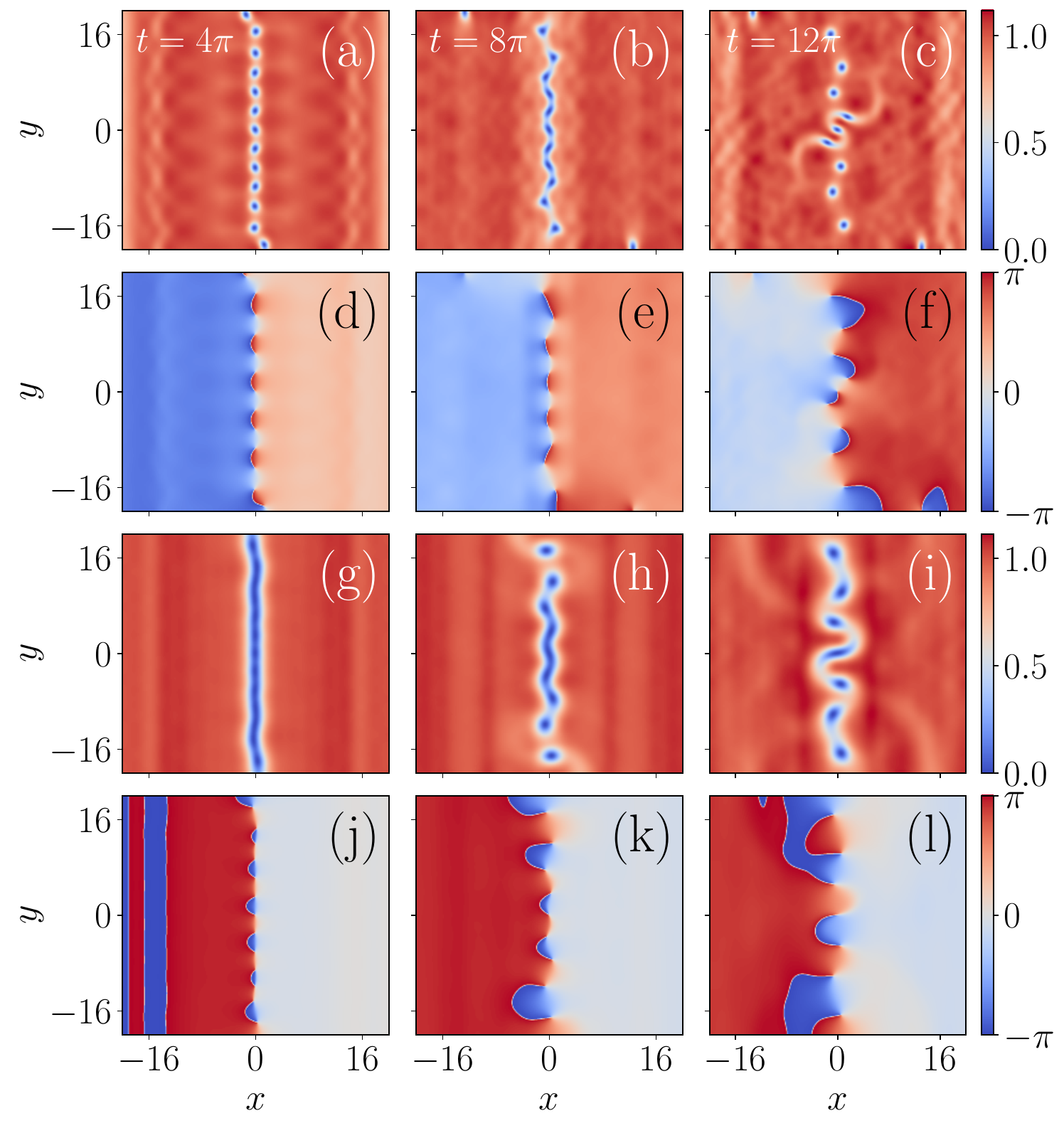}
%\centering\includegraphics[width=\linewidth]{fig02}
\caption{Evolution of a dark soliton subject to snake instability, leading to the formation of a vortex necklace, for an initial modulation amplitude \(A = 0.25\), wavenumber \(k = 1.0\), and chemical potential \(\mu = 1.0\) in the absence of a reservoir (\(m_R = 0\)). For the repulsive three-body interaction case (\(\chi = 1.0\)), (a)--(c) show the density evolution and (d)--(f) the associated phase profiles, where discrete \(2\pi\) phase jumps and the appearance of a vortex--antivortex pair in (f) confirm the topological defect structure. For the attractive three-body interaction case (\(\chi = -0.2\)), (g)--(i) depict the density evolution and (j)--(l) the corresponding phase profiles.}
\label{fig:den-3body}
\end{figure}

Figure~\ref{fig:den-3body} illustrates the dynamics of a dark soliton undergoing snake instability, leading to the emergence of a well-defined vortex necklace. %
This pattern formation results from the interplay between the initial soliton profile and the repulsive three-body interaction (\(\chi = 1\)), with the instability seeded by a small modulation amplitude (\(A = 0.25\)) in the absence of reservoir effects (\(m_R = 0\)). The temporal density evolution shown in Figs. \ref{fig:den-3body}(a)-(c) captures the progressive breakup of the soliton into discrete topological defects, demonstrating the transition from a one-dimensional soliton stripe to a two-dimensional array of vortices driven by the snake instability.

The corresponding phase profiles in Figs. \ref{fig:den-3body}(d)-(f) provide clear topological evidence of the emerging vortex states. The characteristic \(2\pi\) phase windings confirm the presence of quantized vortices. At the same time, the structure in Fig. \ref{fig:den-3body}(f) distinctly reveals the formation of a vortex-antivortex pair, a hallmark of topological excitation associated with the decay of dark solitons in confined geometries~\cite{Lagoudakis2008, Keeling2008, Roumpos2010}. 

Overall, we find that repulsive three-body interactions play a crucial role in determining the instability pathway and the resulting vortex configuration. Rather than producing disordered or turbulent dynamics, the higher-order interactions guide the system towards an organized topological arrangement manifested as a vortex necklace. This underscores the importance of beyond-two-body interactions in dictating nonlinear dynamics and pattern formation in quantum fluids, and suggests promising avenues for controlled vortex engineering in polariton systems.

\subsection{Weak-pumping regime: Repulsive vs attractive interactions}

\paragraph{Repulsive three-body interactions.}

In the weak-pumping regime, binary interactions predominantly drive vortex formation, with the reservoir density adjusting adiabatically to variations in the condensate density. %
%%%%%%%%%%%%%%%%%%%%%%%%%%%%%%%%%%%%%%%%%%%%%%%%%%%
\begin{figure}[!ht]
\centering\includegraphics[width=\linewidth]{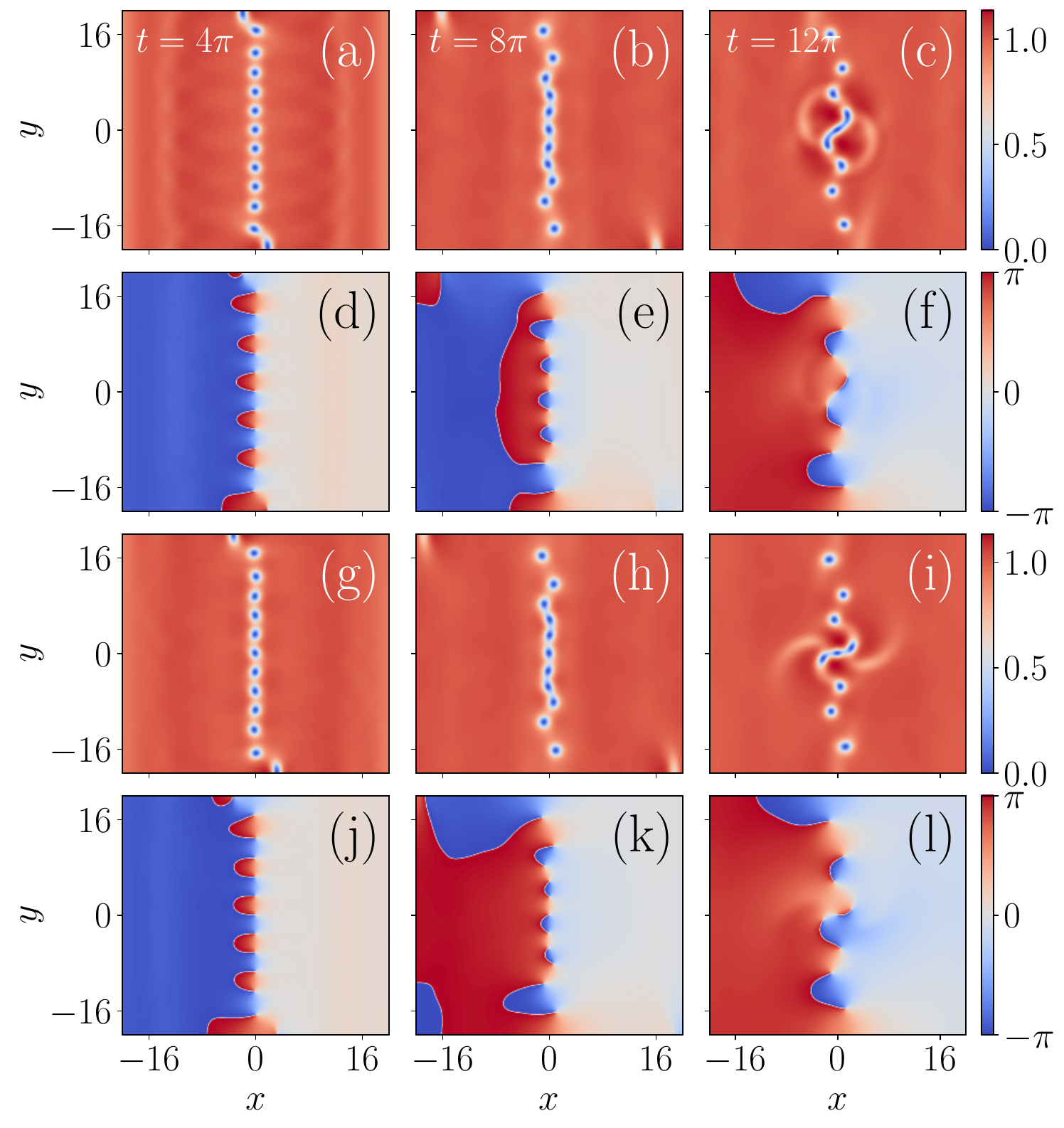}
\caption{Comparative spatiotemporal evolution of dark solitons undergoing snake instability for distinct initial perturbation amplitudes in the presence of repulsive three-body interactions ($\chi = 1.0$) and reservoir coupling, evaluated at wavenumber $k = 1$ and chemical potential $\mu = 1$. Panels (a)--(c) display the density evolution and panels (d)--(f) the corresponding phase profiles for an initial amplitude $A = 0.25$ at times $t = 4\pi$, $8\pi$, and $12\pi$, respectively. Panels (g)--(i) depict the density evolution and panels (j)--(l) the associated phase profiles for $A = 0.5$ at identical time steps. Density distributions illustrate the progressive breakup of the dark soliton into vortex–antivortex pairs, whereas the phase maps delineate characteristic phase windings and reservoir-induced damping. Fixed system parameters are set to $g_R = 2$, $\gamma_C = 3$, $\gamma_R = 15$, and $R = 0.5$.}
\label{fig:weak_1}
\end{figure}%
%%%%%%%%%%%%%%%%%%%%%%%%%%%%%%%%%%%%%%%%%%%%%%%%%%%
However, the introduction of repulsive three-body interactions ($\chi = 1.0$) significantly modifies the underlying condensate dynamics, inducing pronounced changes in both the onset and subsequent development of the transverse snake instability.

Figure~\ref{fig:weak_1} illustrates the comparative evolution of dark solitons undergoing snake instability across two distinct initial instability amplitudes ($A = 0.25$ and $A = 0.5$) under repulsive three-body interactions ($\chi = 1$) and active reservoir coupling. The density profiles indicate that elevating the initial instability amplitude substantially accelerates the disintegration of the dark soliton. For the lower amplitude case ($A = 0.25$, shown in panels (a)--(c)), the soliton exhibits a relatively gradual transverse modulation, resulting in a delayed onset of the snake instability and a constrained nucleation of vortex– antivortex pairs. Conversely, for the higher amplitude regime ($A = 0.5$, shown in panels (g)--(i)), the instability develops at an accelerated rate, yielding a pronounced and earlier structural fragmentation of the soliton. This rapid decay drives a prolific generation of vortex antivortex pairs, signifying enhanced growth rates of unstable transverse modes. The ensuing dynamics demonstrate that these vortex pairs propagate in opposite transverse directions, a hallmark of the signature of the snake instability, while their increasing spatial density signals a transition from a coherent solitonic framework to a disordered, turbulent hydrodynamic regime.

Further insight into the system dynamics is provided by the corresponding phase profiles [shown in Fig.~\ref{fig:weak_1}(d)--(f) and (j)--(l)]. In both instances, the phase distributions exhibit well-defined $2\pi$ phase windings localized at the vortex cores, confirming the generation of topological defects following soliton decay. For $A = 0.25$ [shown in Fig.~\ref{fig:weak_1}(d)--(f)], the phase structure evolves smoothly, reflecting the subdued nucleation and separation kinetics of the vortex pairs. In contrast, for $A = 0.5$ [shown in Fig.~\ref{fig:weak_1}(j)--(l)], the phase field undergoes rapid fragmentation at earlier evolutionary stages, consistent with the heightened vortex proliferation observed in the density profiles. Furthermore, the phase evolution captures the dissipative influence of the reservoir coupling, introducing viscous-like damping into the system. This phenomenon manifests as a gradual smoothing of phase gradients, which subsequently alters both the trajectories and operational lifetimes of the emergent vortex pairs. Consequently, the delicate interplay between nonlinear interactions and reservoir-driven dissipation dictates not only the critical threshold for instability onset, but also the long-term dynamics and persistence of the topological structures.

As analytically indicated by Eq.~\eqref{life-time-weak}, the lifetime of the resulting vortex configuration exhibits explicit dependence on the parameters $\gamma_C$, $\gamma_R$, and $A$. 
%To complement these findings, the long-time dynamics are animated in the Supplementary Material~\cite{video}, capturing the continuous condensate density evolution from $t = 0$ to $t = 60\pi$. This visual evidence highlights the complete lifecycle of the snake instability, from initial soliton bending and vortex nucleation to propagation and eventual long-term decay, thereby serving as a dynamic extension to the discrete snapshots presented in Fig.~\ref{fig:weak_1}.

To further elucidate the role of active reservoir coupling within this repulsive regime, we examine how varying the stimulated scattering rate influences the stability and lifetime of the emergent vortex structures. 

%%%%%%%%%%%%%%%%%%%%%%%%%%%%%%%%%%%
\begin{figure}[!ht]
\centering\includegraphics[width=\linewidth]{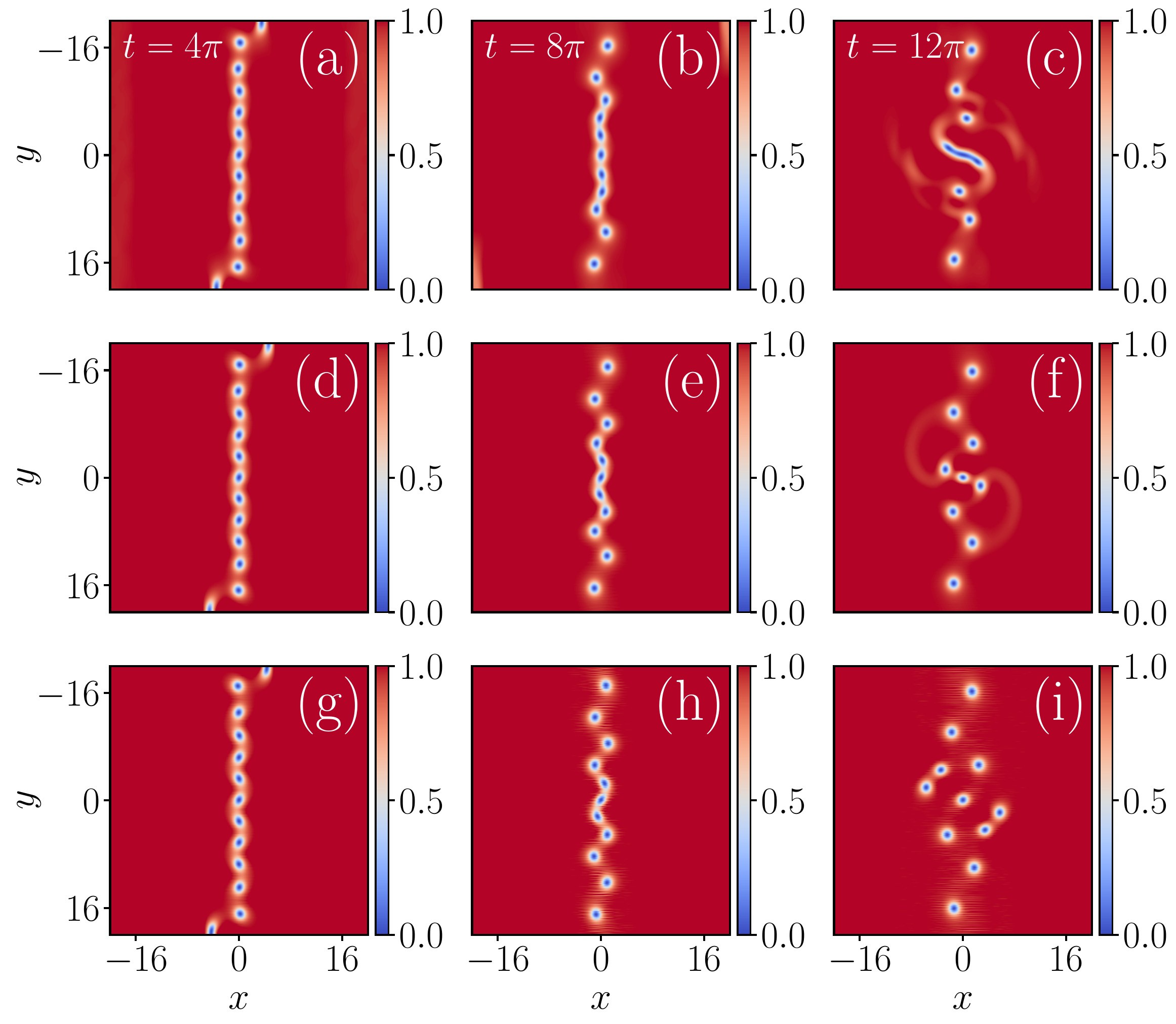}
\caption{Time evolution of density profiles illustrating the snake instability of a dark soliton in the presence of repulsive three-body interactions (\(\chi = 1.0\)), for an initial perturbation amplitude \(A = 0.5\), wavenumber \(k = 1\), and chemical potential \(\mu = 1\) under weak pumping conditions. Panels (a)--(i) illustrate vortex formation influenced by reservoir coupling at varying stimulated scattering rates: (a)--(c) \(R = 1\), (d)--(f) \(R = 5\), and (g)--(i) \(R = 10\). Remaining parameters are fixed as \(g_R = 2\), \(\gamma_C = 3\), and \(\gamma_R = 15\).}
\label{fig:weak_perturb}
\end{figure}  
%%%%%%%%%%%%%%%%%%%%%%%%%%%%%%%%%%%

As shown in Fig.~\ref{fig:weak_perturb}, the snake instability develops cleanly devoid of boundary interference, confirming that dark solitons undergo this transverse modulation while nucleating stable vortices. 

For weak scattering (\(R = 1\) at \(A = 0.5\)), the characteristic lifetime \(\tau\) is established relative to the reference timescale \(\tau_0\) [cf. Fig.~\ref{fig:weak_perturb}(a)--(c)]. Increasing the reservoir scattering rate accelerates the decay dynamics, leading to rapid vortex destabilization: at \(R = 5\), the lifetime decreases substantially as boundary effects become more pronounced [cf. Fig.~\ref{fig:weak_perturb}(d)--(f)], and at \(R = 10\), the topological structures undergo severe degradation and dissolve [cf. Fig.~\ref{fig:weak_perturb}(g)--(i)]. This demonstrates that enhanced reservoir scattering in the repulsive regime exacerbates structural breakdown rather than stabilizing the condensate.

Having established the role of repulsive  three-body interactions and reservoir coupling in the exciton-polariton condensate, we now extend our analysis to contrasting dynamical regimes. Incorporating these competing nonlinearities fundamentally alters the balance of forces within the system, driving distinct instability pathways, altered vortex dynamics, and contrasting hydrodynamic flow behaviors relative to the repulsive case.

\paragraph{Attractive three-body interactions.}
We now investigate the effect of attractive three-body interactions (\(\chi < 0\)) on the dynamical stability of dark solitons under weak pumping. This scenario contrasts sharply with the repulsive regime, highlighting how the sign of the three-body nonlinearity governs vortex formation and persistence.

\begin{figure}[!ht]
\centering
\includegraphics[width=\linewidth]{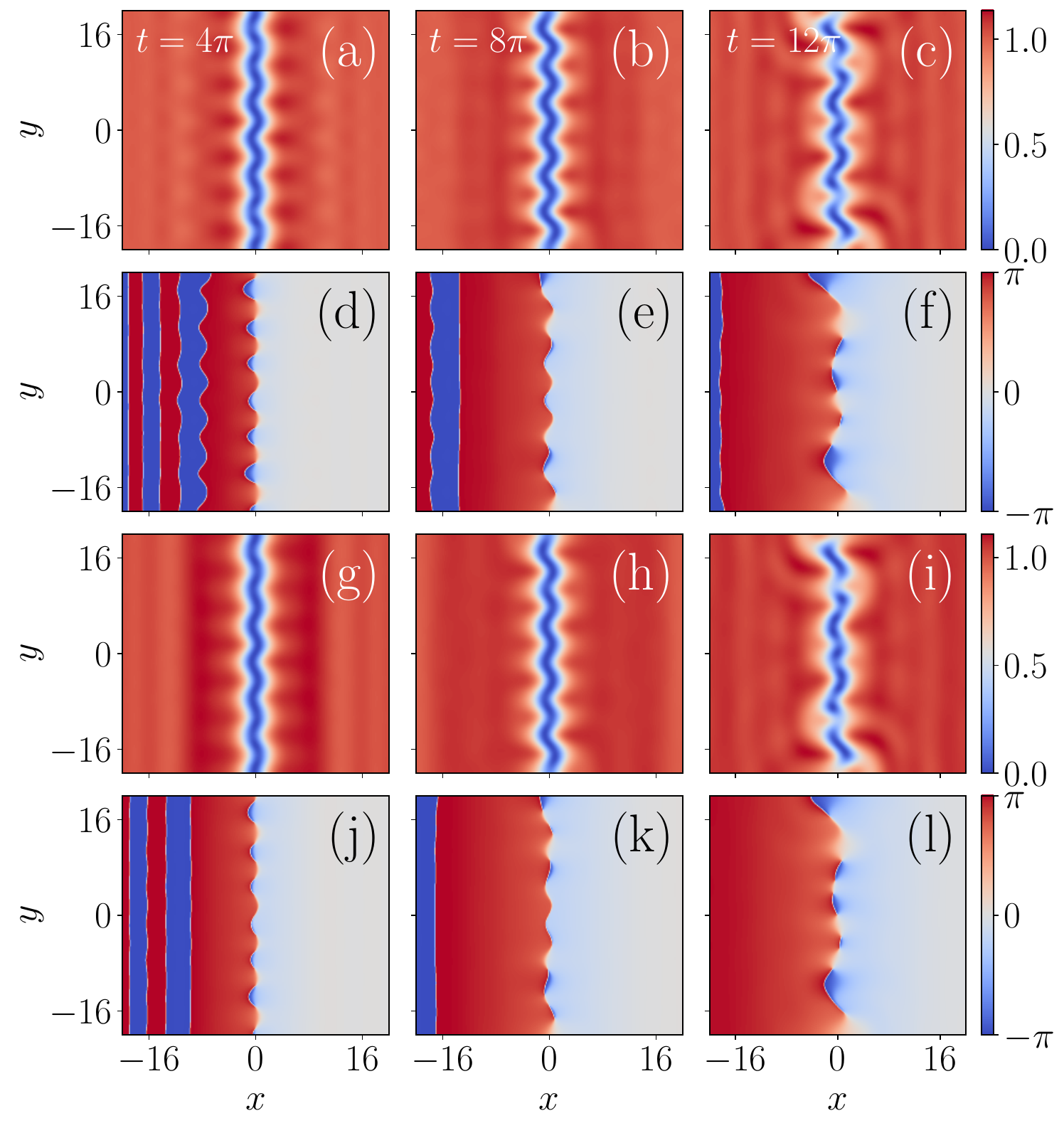}
\caption{Comparative evolution of dark solitons undergoing snake instability for different chemical potentials in the presence of attractive three-body interactions (\(\chi = -0.2\)) and reservoir coupling. The top two rows display the density evolution [(a)--(c)] and corresponding phase profiles [(d)--(f)] for an initial chemical potential of \(\mu = 0.25\). The bottom two rows show the density evolution [(g)--(i)] and phase profiles [(j)--(l)] for \(\mu = 0.5\). In both cases, the wavenumber is \(k = 1\), and the instability amplitude is \(A = 0.25\). Remaining parameters are fixed as \(g_R = 2\), \(\gamma_C = 3\), \(\gamma_R = 15\), and \(R = 0.5\).}
\label{fig:6}
\end{figure}

Figure~\ref{fig:6} illustrates the condensate dynamics in the presence of an attractive three-body interaction (\(\chi = -0.2\)) combined with repulsive binary interactions, evaluated at a fixed instability amplitude of \(A = 0.25\).  
%%%%%%%%%%%%%%%%%%%%%%%%%%%%%%%%%%%%%%%%%%%%%%%%%%%%%%
%%%%%%%%%%%%%%%%%%%%%%%%%%%%%%%%%%%%%%%%%%%%%%%%____
As shown in Fig.~\ref{fig:6}, although vortices are initially nucleated along the dark soliton, they fail to maintain coherent structures over time. The resulting vortex necklace is inherently unstable and gradually dissipates in stark contrast to the repulsive case, where stable configurations persist. This rapid breakdown leads to a highly transient and unstructured flow within the condensate, confirming that attractive interactions introduce a strong tendency toward decay and inhibit long-lived topological defects.

\begin{figure}[!ht]
\centering
\includegraphics[width=\linewidth]{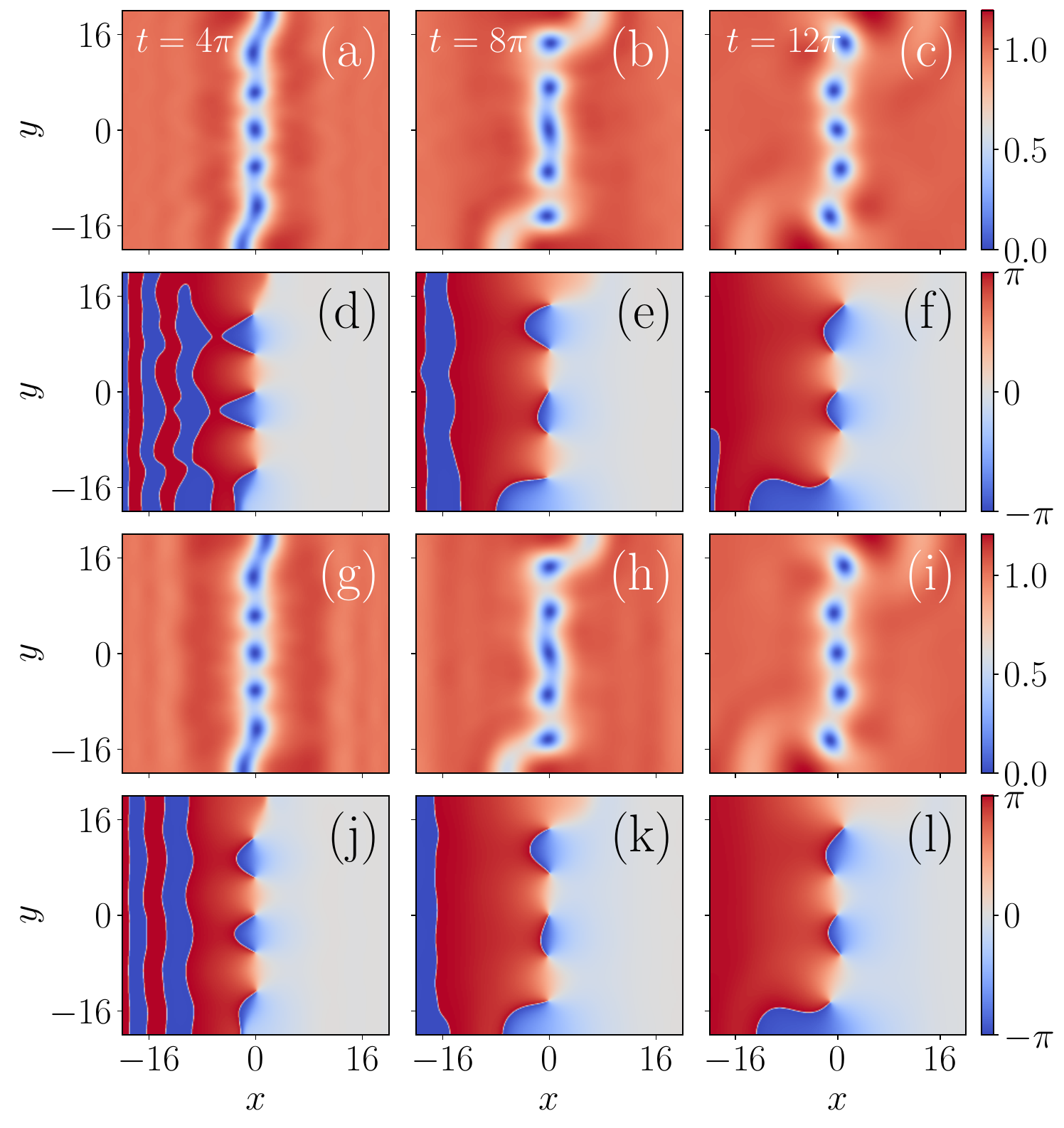}
\caption{Comparative evolution of dark solitons undergoing snake instability for different chemical potentials in the presence of attractive three-body interactions (\(\chi = -0.2\)) and reservoir coupling. The top two rows display the density evolution [(a)--(c)] and phase profiles [(d)--(f)] for \(\mu = 0.25\). The bottom two rows present results for \(\mu = 0.5\) [(g)--(i) for density, (j)--(l) for phase]. In both cases, the wavenumber is \(k = 0.5\), and the instability amplitude is \(A = 0.5\). Remaining parameters are fixed as \(g_R = 2\), \(\gamma_C = 3\), \(\gamma_R = 15\), and \(R = 0.5\).}
\label{fig:7}
\end{figure}

Increasing the instability amplitude further amplifies this behavior, as demonstrated in Fig.~\ref{fig:7} for \(A = 0.5\) and \(k = 0.5\). The number of unstable vortices scales with larger amplitudes, reflecting the system's heightened susceptibility to dynamical instabilities under attractive nonlinearities. Unlike the repulsive regime, where each pronounced bending of the dark soliton typically yields a single stable vortex, the attractive interaction produces multiple vortices per instability cycle. However, these structures remain short-lived and fail to establish persistence, culminating in an increasingly disordered and chaotic pattern. As corroborated by our asymptotic analysis and Eq.~\eqref{life-time-weak}, the vortex lifetime in this regime continues to be governed by \(\gamma_C\), \(\gamma_R\), and \(A\).  
%%%%%%%%%%%%%%%%%%%%%%%%%%%%%%%%%%%%%%%%%%%
%%%%%%%%%%%%%%%%%%%%%%%%%%%%%%%%%%%%%%%%%%%%%%%%%%%%%%%
As seen in Fig.~\ref{fig:7}, another striking consequence of attractive three-body interactions is the spatial broadening of the snake instability along the soliton trajectory. While the repulsive regime confines the instability to a relatively narrow channel, attractive interactions induce larger transverse deformations of the soliton. This broadening facilitates the nucleation of a greater number of vortex--antivortex pairs; however, their intrinsic instability ensures that these structures remain transient. Overall, attractive interactions enhance both the amplitude and spatial extent of soliton bending, promoting more irregular and turbulent condensate dynamics.

\begin{figure}[!ht]
\centering
\includegraphics[width=\linewidth]{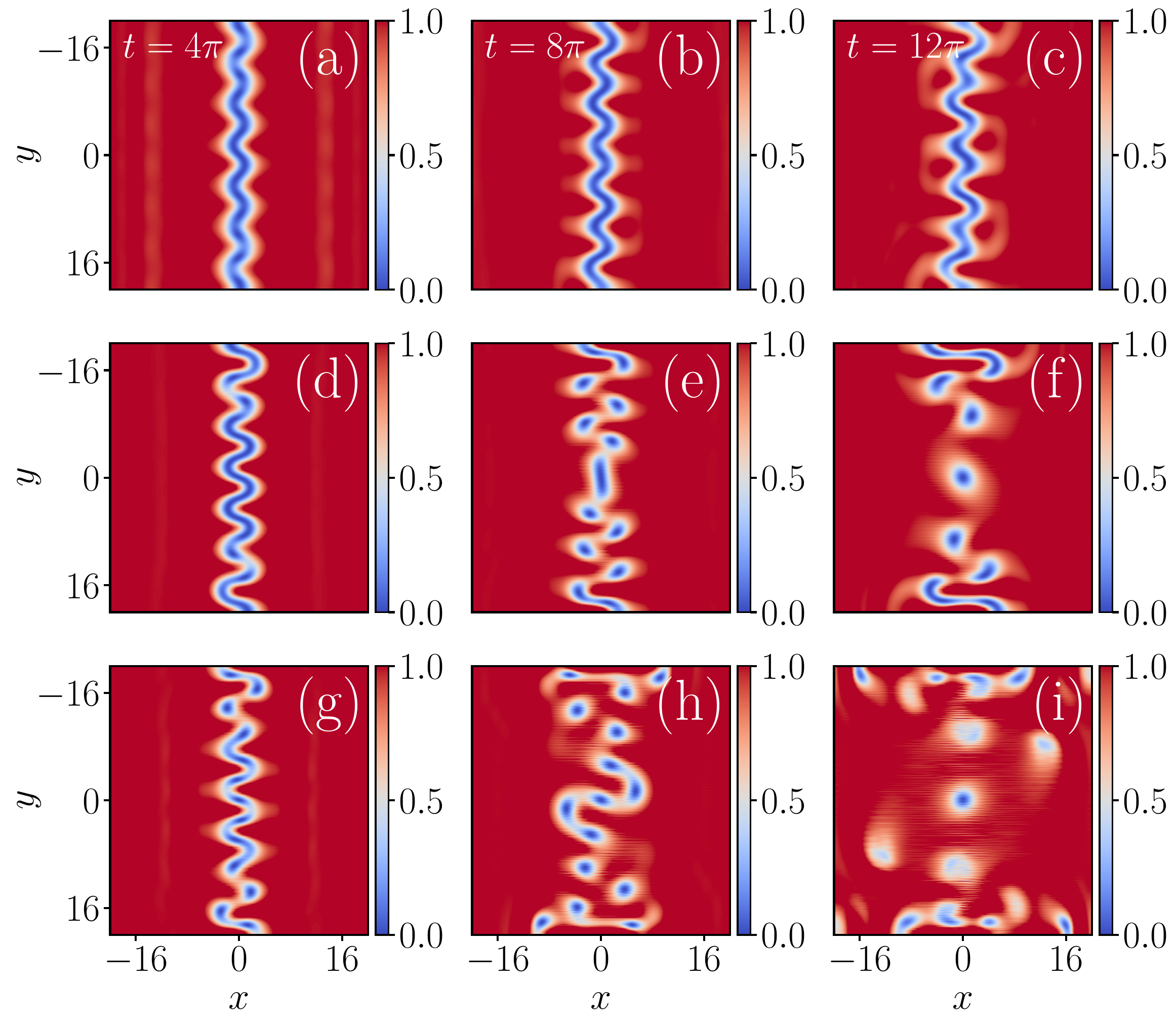}
\caption{Time evolution of density profiles illustrating the snake instability of a dark soliton in the presence of attractive three-body interactions (\(\chi = -0.2\)), for an initial perturbation amplitude \(A = 0.5\), wavenumber \(k = 1.0\), and chemical potential \(\mu = 1.0\) under weak pumping conditions. Panels (a)--(i) illustrate vortex formation influenced by reservoir coupling at varying stimulated scattering rates: \(R = 1\) [(a)--(c)], \(R = 5\) [(d)--(f)], and \(R = 10\) [(g)--(i)]. Remaining parameters are fixed as \(g_R = 2\), \(\gamma_C = 3\), and \(\gamma_R = 15\).}
\label{fig:weak_perturb:1}
\end{figure}

Furthermore, the snake instability in the attractive domain exhibits heightened sensitivity to boundary effects, which play a decisive role in disrupting the coherence of the vortex configurations, as evidenced in Fig.~\ref{fig:weak_perturb:1}. 

%%%%%%%%%%%%%%%%%%%%%%%%%%%%%%%%%
%%%%%%%%%%%%%%%%%%%%%%%%%%%%%%%%%%%

As depicted in Fig.~\ref{fig:weak_perturb:1}, at a low scattering rate of \(R = 1\) [cf. Fig.~\ref{fig:weak_perturb:1}(a)--(c)], the system evolves relatively slowly. The initial configuration displays a dark soliton with slight distortions driven by attractive interactions [panel (a)], leading to the emergence of small, localized vortices at the soliton edges that remain transiently stable before boundary reflections trigger early-stage instability. At a moderate scattering rate of \(R = 5\) [cf. Fig.~\ref{fig:weak_perturb:1}(d)--(f)], boundary effects become more pronounced and accelerate the transition toward global instability. While the initial soliton profile remains identifiable [cf. Fig.~\ref{fig:weak_perturb:1}(d)], vortex formation grows chaotic, and boundary-adjacent vortices interact vigorously to accelerate their mutual destabilization [cf. Fig.~\ref{fig:weak_perturb:1}(e)], driving the system into an irregular turbulent state until structural integrity is lost [cf. Fig.~\ref{fig:weak_perturb:1}(f)]. At the highest scattering rate of \(R = 10\) [cf. Fig.~\ref{fig:weak_perturb:1}(g)--(i)], degradation is severe: the soliton experiences drastic deformation and fragmentation [cf. Fig.~\ref{fig:weak_perturb:1}(g)], while highly mobile vortices collide, fragment, and rapidly dissolve [cf. Fig.~\ref{fig:weak_perturb:1}(h)]. Ultimately, strong scattering and boundary reflections completely overwhelm and dissolve the topological structures, leaving the system in a strongly disordered state [cf. Fig.~\ref{fig:weak_perturb:1}(i)].

In contrast to the repulsive domain, where vortex structures exhibit robust stability, the attractive regime is marked by rapid vortex destabilization. This instability is heavily influenced by boundary effects arising from the interplay between attractive interactions and reservoir coupling. Consequently, as the scattering rate increases, the vortices undergo progressive destabilization, eventually breaking down and dissolving completely.

\subsection{Strong-pumping regime: Repulsive vs attractive interactions}

\begin{figure}[!ht]
\centering
\includegraphics[width=\linewidth]{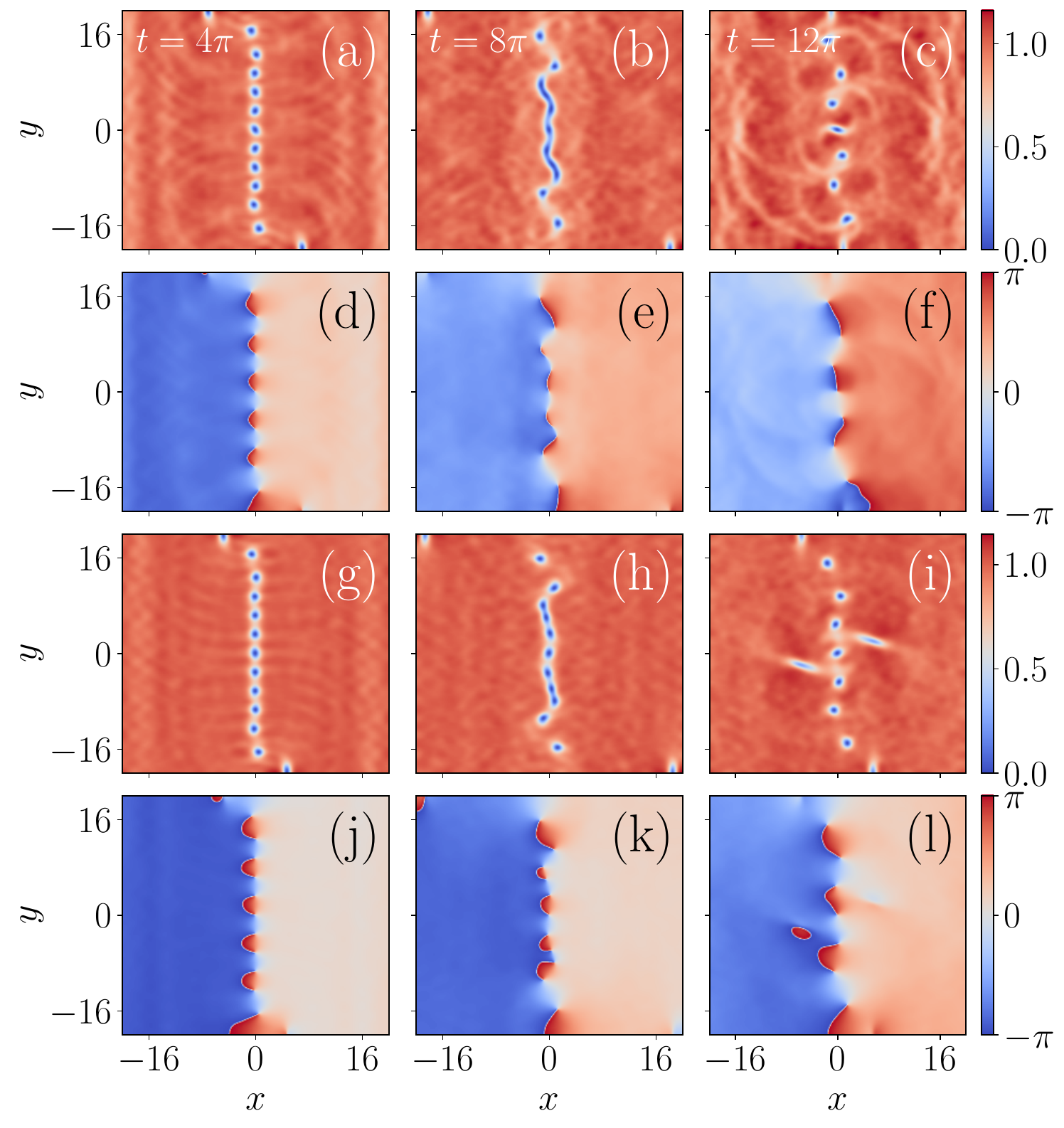}
\caption{Evolution of a dark soliton undergoing snake instability in the presence of repulsive three-body interactions ($\chi = 1$) and reservoir coupling. The initial instability amplitude is $A=0.9$, and the scattering rate is $R=0.08$. Panels (a-c) show the development of vortices leading to snake instability, and panels (d-f) show the corresponding phases. This evolution ultimately results in the breakdown of the condensate due to strong reservoir pumping at $\mu = 0.5$. Panels (g-i) show the vortex dynamics for a higher pump strength, $\mu = 1.0$, while the corresponding phases are shown in panels (j-l). Other parameters are $g_{R}= 2$, $ \gamma_{C}=0.0075$, and $\gamma_{R}= 0.07$.}
\label{fig:den3}
\end{figure}

\paragraph{Repulsive three-body interactions.}
In the strong-pumping regime, at a higher instability amplitude of \(A = 0.9\), the system supports the formation of a stable vortex necklace, as shown in Fig.~\ref{fig:den3}. 
%%%%%%%%%%%%%%%%%%%%%%%%%%%%%%%%%%%%%%%%%%%%
%%%%%%%%%%%%%%%%%%%%%%%%%%%%%%%%%%%%%%%%%%%%
Unlike the transient vortex-antivortex pairs observed at lower amplitudes, the vortex necklace constitutes a highly structured and robust configuration. These results highlight the intricate interplay between pumping strength, three-body interactions, reservoir-induced damping, and instability amplitude in determining vortex stability in exciton-polariton condensates. The nonlocal coupling between the reservoir and the condensate polaritons ensures the persistence of the vortex over a full cycle, in agreement with the asymptotic description under strong pumping for both repulsive and attractive three-body interactions, as discussed in Sec.~\ref{sec:strong}.

\begin{figure}[!ht]
\centering
\includegraphics[width=\linewidth]{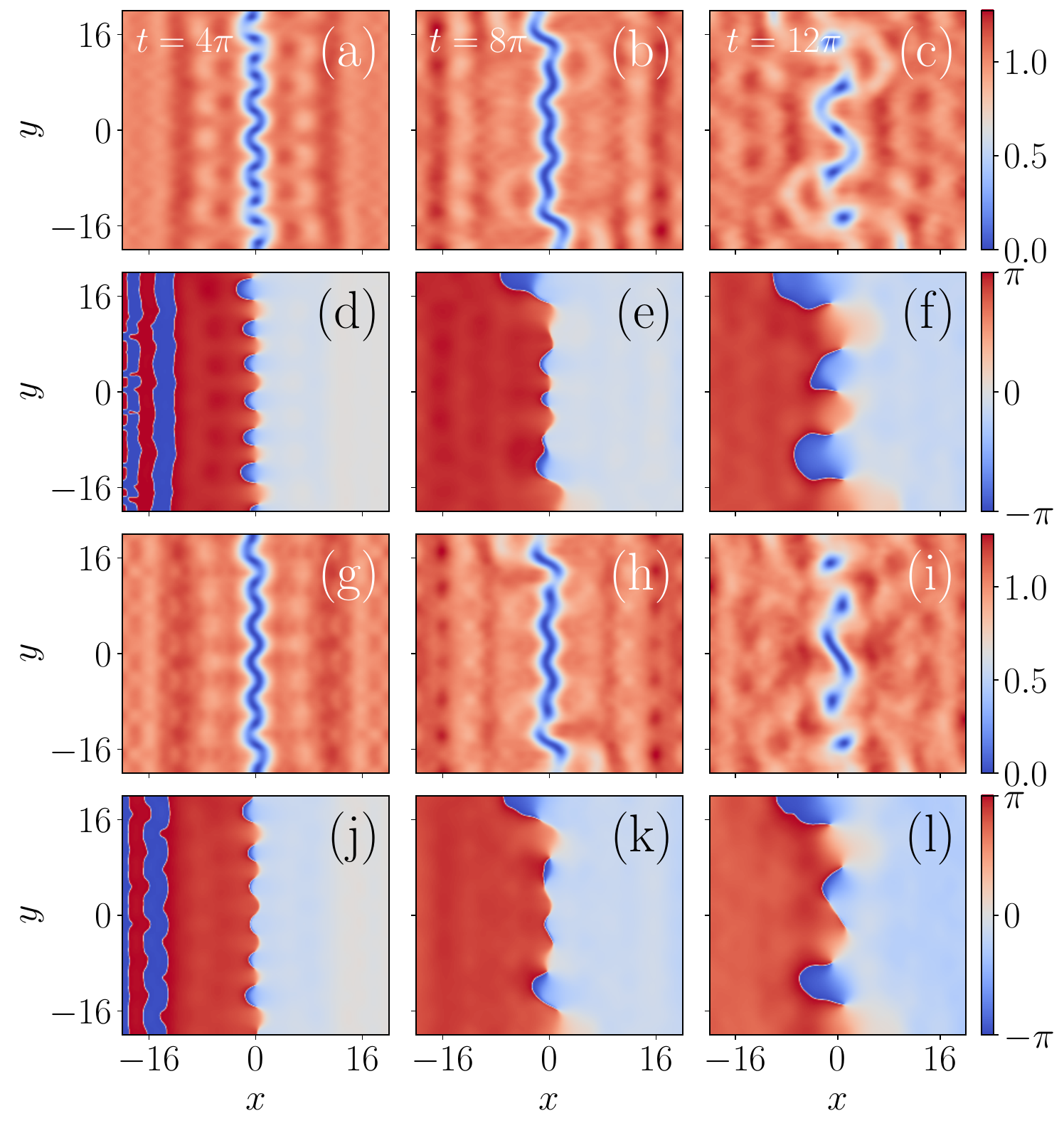}
\caption{Comparative evolution of the dark solitons undergoing snake instability for different chemical potentials in the presence of attractive three-body interaction $(\chi=-0.2)$ and the reservoir coupling. The top two rows correspond to an initial chemical potential of $\mu=0.5$: panels (a)-(c) show the density evolution, while (d)-(f) display the corresponding phase profiles. The bottom two rows present the results of $\mu=1.0$: panels (g)-(i) depict the density evolution, and (j)-(l) show the associated phase profiles. In both cases, the wave number is $k=1.0$, and the instability amplitude is $A=0.9$. The remaining parameters are fixed as $g_{R}= 2$, $\gamma_{C}=0.0075$, $\gamma_{R}= 0.07$ and $R=0.08$.}
\label{fig:8}
\end{figure}

\paragraph{Attractive three-body interactions.}
For strong pumping, we observe a complete distortion of the snake instability, as shown in Fig.~\ref{fig:8}. 
%%%%%%%%%%%%%%%%%%%%%%%%%%%%%%%%%%%%%%%%%%%%%%%%%%%%%%%%
%%%%%%%%%%%%%%%%%%%%%%%%%%%%%%%%%%%%%%%%%%%%%%%%%%%%%%%%
Unlike the repulsive domain, where the snake instability results in the stable formation of a vortex necklace, the attractive three-body interaction leads to a more pronounced and rapid breakdown of the vortex structure. This results in a chaotic and less stable configuration of vortices, with the instability further distorting the soliton profile. The complete distortion of the instability under strong pumping contrasts sharply with the behavior seen in the repulsive regime, where vortex structures remain relatively stable.
Overall, we find that the attractive interaction leads to greater instability, faster breakdown of vortex structures, and more complex, transient vortex patterns compared to the repulsive case.

\section{Lifetime of the vortices}
\label{sec:LifetimeVortices}
The role of the reservoir in determining the lifetime of the vortex structures can be understood from the effective gain--loss balance experienced by the condensate. In the open-dissipative Gross--Pitaevskii description, the reservoir provides both a source of gain and a spatially dependent contribution to the condensate dynamics. Since the reservoir density is coupled to the condensate density, the relevant decay rate is therefore more appropriately characterized by a condensate-density-weighted reservoir contribution rather than by the bare condensate loss rate. This provides a useful measure of the effective lifetime of the vortex structures in the driven-dissipative system as,
\begin{align}
\Gamma_{\mathrm{eff}}(t) ={\gamma}_{C}-{R}\left\langle m_R \right\rangle{|\psi|^2}.
\label{eq:Lifetime}
\end{align}
This quantity measures the instantaneous balance between condensate loss and reservoir-induced replenishment and therefore captures the temporal evolution of the dissipative environment in which the vortex structures evolve.

\begin{figure}[!htbp]
    \centering
    \includegraphics[width=\linewidth]{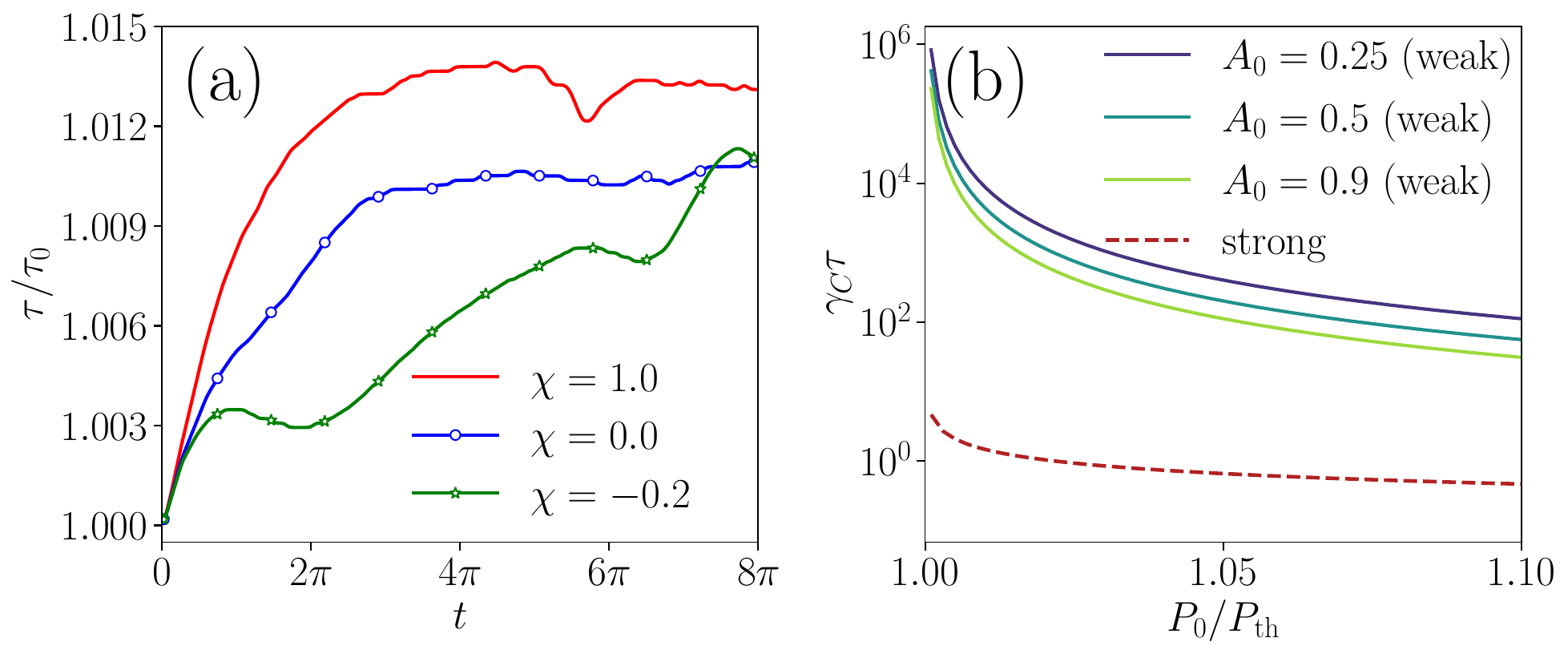}
    \caption{(a) Effective lifetime ($\tau/\tau_0$) as a function of time $t$ calculated from the condensate-density-weighted reservoir contribution for repulsive ($\chi=1.0$), baseline binary ($\chi=0.0$), and attractive ($\chi=-0.2$) three-body interactions with other parameters fixed at $g_R=2$, $\gamma_C = 0.007$, and $\gamma_R = 0.07$. (b) Normalized lifetime $\gamma_C \tau$ as a function of relative pump power $P_0/P_{\mathrm{th}}$ for various initial perturbation amplitudes ($A_0 = 0.25, 0.5, 0.9$) in the weak- and strong-pumping regimes. The parameters for the weak-pumping regime in (b) are $g_R = 2$, $\gamma_C = 3$, $\gamma_R = 15$, and $R = 5.0$, while for the strong-pumping regime, $g_R = 2$, $\gamma_C = 3$, $\gamma_R = 15$, and $R = 0.08$.}
    \label{fig:lifetime_combined}
\end{figure}

Figure~\ref{fig:lifetime_combined}(a) shows the effective lifetime for repulsive and attractive three-body interactions, corresponding to $\chi=1$ and $\chi=-0.2$, respectively. 
In both cases, $\tau_{\mathrm{eff}}$ increases gradually with time, indicating that the reservoir-mediated gain becomes progressively more effective in compensating condensate losses as the system evolves. The increase is systematically stronger for the repulsive three-body interaction. Thus, the repulsive three-body nonlinearity provides a more favorable dissipative environment for sustaining the vortex structures, whereas the attractive interaction leads to a comparatively shorter effective lifetime.

The dependence on the pumping conditions is further illustrated by the decay rates shown in Fig.~\ref{fig:lifetime_combined}(b), where different instability amplitudes, $A_0=0.25$, $0.5$, and $0.9$, are considered in the weak- and strong-pumping regimes. The resulting decay rates demonstrate that the lifetime is controlled by the interplay between the reservoir-induced gain, condensate loss, pumping strength, and nonlinear instability. In particular, the stronger pumping regime modifies the gain--loss balance and consequently changes the rate at which the instability develops.

\begin{figure}[!ht]
    \centering
    \includegraphics[width=\linewidth]{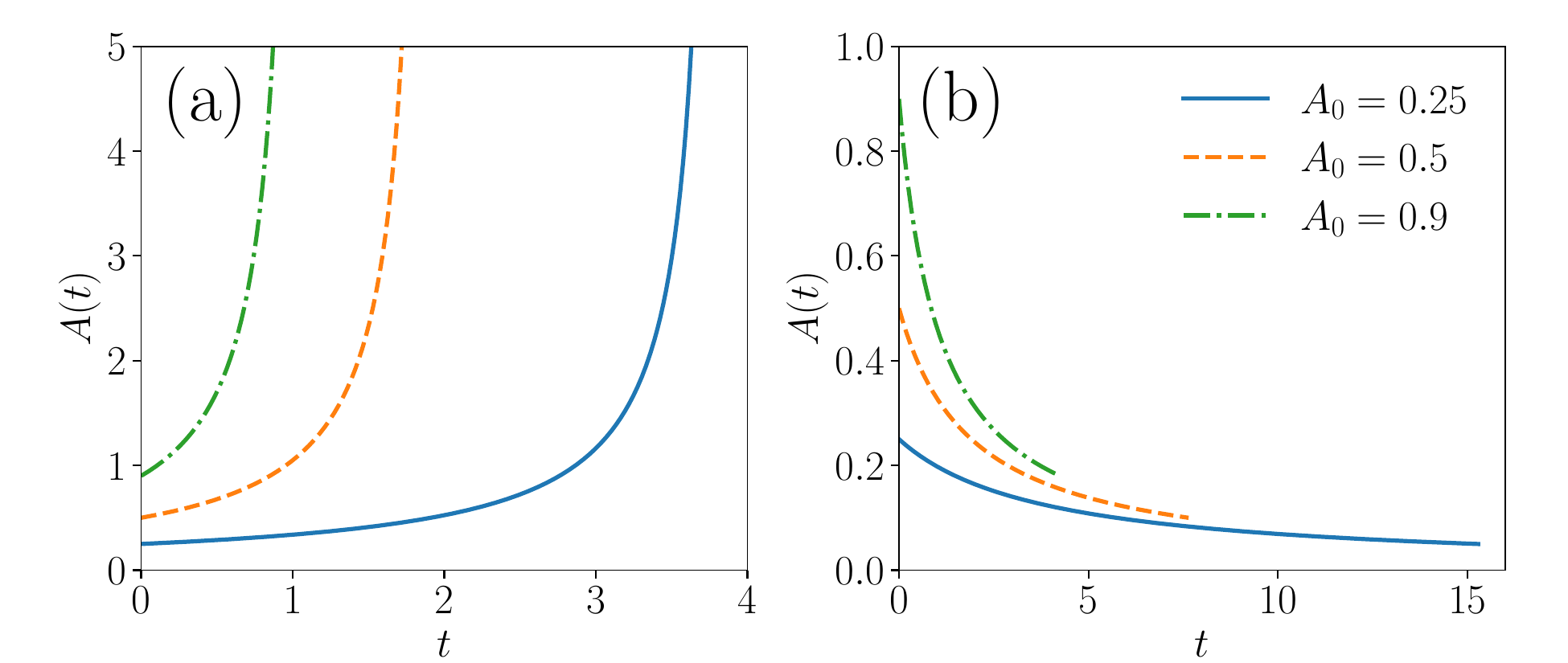}
    \caption{Time evolution of the instability amplitude $A(t)$ for various initial amplitudes $A_0$ under reservoir coupling: (a) repulsive three-body interaction ($\chi = 1.0$) with parameters identical to Fig.~\ref{fig:den3}, and (b) attractive three-body interaction ($\chi = -0.2$) with parameters identical to Fig.~\ref{fig:8}.}
    \label{fig:amp-pump}
\end{figure}

This difference becomes particularly relevant when combined with the dynamics of the instability amplitude. Figure~\ref{fig:amp-pump} shows the evolution of the instability amplitude $A$ for different initial amplitudes $A_0$ in the presence of reservoir coupling. For $\chi=1$, the reservoir-induced modification of the instability remains sufficiently weak to preserve the vortex structures over an extended time interval. In contrast, for $\chi=-0.2$, the instability is more strongly amplified near the condensate boundary, facilitating the subsequent deformation and disintegration of the vortex structures. The reservoir therefore does not merely introduce an overall damping of the condensate; its dynamical coupling to the instability can selectively enhance or suppress the growth of spatial modes. %

Taken together, these results provide a dynamical explanation for the enhanced robustness of vortices in the repulsive three-body interaction regime. The reservoir coupling continuously modifies both the effective lifetime and the growth of the instability, while the sign of the three-body interaction determines how strongly the resulting density perturbations evolve. For $\chi>0$, the repulsive nonlinear contribution suppresses the growth of boundary distortions and, together with the reservoir-mediated gain, favors long-lived vortex configurations. For $\chi<0$, the attractive three-body interaction enhances the nonlinear deformation of the condensate and promotes the boundary-driven growth of the instability, ultimately shortening the vortex lifetime. In other words, for attractive three-body interactions, the instability amplitude rapidly decays with time, thereby suppressing vortex formation in the condensates. In contrast, repulsive three-body interactions lead to a pronounced growth of the instability amplitude at long times, which promotes the nucleation and subsequent formation of vortices. This behavior is consistent with the central finding that repulsive three-body interactions stabilize the vortex structures, whereas attractive three-body interactions facilitate their eventual disintegration.

\section{Summary and Conclusion}
\label{sec:con}
In this paper, we have developed a comprehensive study of the dynamics and stability of exciton-polaritons within the framework of the open-dissipative Gross-Pitaevskii equation, incorporating both binary and short-range three-body interactions. Using an asymptotic description, we derived equations for the instability amplitude, offering new insights into vortex formation through the snake instability of dark solitons. Our findings reveal that the interplay between binary and three-body interactions plays a crucial role in the stability and behavior of vortices in these systems.

In particular, we found that a repulsive three-body interaction, when combined with binary interactions, promotes the formation of stable vortex-antivortex pairs and robust vortex necklaces. These structures exhibit stable dynamics over extended periods in the repulsive regime. In contrast, the introduction of attractive three-body interactions exacerbates the snake instability, leading to rapid vortex disintegration driven by boundary effects. These boundary effects, modulated by the coupling to the reservoir, destabilize vortices more efficiently in the attractive regime, underscoring the destructive influence of boundary-induced perturbations.

The time evolution of the instability, governed by reservoir effects and varying pumping conditions, further highlights the pronounced contrast between the repulsive and attractive three-body interaction regimes. While repulsive interactions provide a favorable dissipative environment that sustains vortex configurations even under strong pumping,  attractive interactions shorten the effective lifetime and promote rapid structural breakdown. The varying width, number, and persistence of vortices observed across different scenarios are fundamentally determined by the nature of the three-body interactions and the driven-dissipative parameters of the exciton-polariton system.

A detailed formalism of the vortex dynamics for exciton-polariton condensates with three-body interactions will be valuable for guiding future experiments in these systems. Extending this analysis to cover both weak- and strong-pumping regimes has revealed how non-equilibrium phenomena, dissipation, and nonlinear interactions shape topological stability. Natural avenues for future work include exploring vortex-antivortex annihilation processes or the formation of non-trivial topological defects under broader parameter landscapes, which could offer deeper insights into topological quantum fluids and the influence of dissipation on macroscopic quantum order.

\acknowledgments
R.~R. acknowledges financial assistance received from DST-CURIE (Grant No.~DST-CURIE/PG/54/2022) and ANRF (Grant No.~DST-CRG/008153/2023) in the form of sponsored research projects. P.~M. acknowledges financial support from MoE RUSA 2.0 (Bharathidasan University -- Physical Sciences).

\begin{comment}
\section*{Author Declarations}

\subsection*{Conflict of Interest}
\noindent The authors have no conflicts to disclose.

\subsection*{Author Contributions}
\noindent\textbf{Palanivel Raman:} 
Investigation (lead); Software (lead); Formal analysis (lead); Visualization (lead); Writing -- original draft (lead).
\noindent\textbf{Ramaswamy Radha:} 
Conceptualization (equal); Methodology (equal); Supervision (equal); Writing -- review \& editing (equal); Funding acquisition.
\noindent\textbf{Pankaj Kumar Mishra:} 
Conceptualization (equal); Methodology (equal); Supervision (equal); Validation (equal); Writing -- review \& editing (equal).
\noindent\textbf{Paulsamy Muruganandam:} 
Conceptualization (equal); Methodology (equal); Software (supporting); Supervision (equal); Validation (equal); Writing -- review \& editing (equal); Funding acquisition.

\section*{Data Availability Statement}
The data supporting the findings of this work can be generated by solving the equations with the methods provided. Additionally, the data can be obtained from the corresponding author upon reasonable request.
\end{comment}

\appendix

\section{Derivation of the Hydrodynamic Velocity Fields}
\label{appendix:velocity_derivation}

In the hydrodynamic formulation of the Gross-Pitaevskii equation (with $\hbar=1$ and $M=1$), the superfluid current density is expressed as
\begin{align}
    \vec{j}(\vec{r},t) = \frac{1}{2i}\left[\Psi^{*}\nabla \Psi - (\nabla \Psi^{*})\Psi\right] = \operatorname{Im}(\Psi^{*}\nabla\Psi).
\end{align}
Equating this to the hydrodynamic form $\vec{j}(\vec{r},t) = n(\vec{r},t)\vec{v}(\vec{r},t)$, with $n(\vec{r},t) = |\Psi|^2$, the superfluid velocity field $\vec{v} = \nabla S$ is given by
\begin{align}
    \vec{v}(\vec{r},t) = \frac{\operatorname{Im}(\Psi^{*}\nabla\Psi)}{|\Psi|^{2}}.
\end{align}
Substituting the perturbed dark-soliton ansatz $\Psi(x,y,t) = \sqrt{n_0} f(u,y,t)$ with $u = x/\xi$, the complex profile function reads
\begin{align}
    f(u,y,t) = \tanh u + i A(t) \operatorname{sech} u \sin(ky).
\end{align}
The corresponding spatial derivatives are
\begin{subequations}
\begin{align}
    \partial_x f &= \frac{1}{\xi} \partial_u f = \frac{1}{\xi}\left[\operatorname{sech}^{2} u - i A(t) \operatorname{sech} u \tanh u \sin(ky)\right], \\
    \partial_y f &= i k A(t) \operatorname{sech} u \cos(ky).
\end{align}
\end{subequations}
The density denominator is given by
\begin{align}
    |f|^{2} = \tanh^{2} u + A^{2}(t)\operatorname{sech}^{2} u \sin^{2}(ky).
\end{align}
In the linear instability regime ($A(t) \ll 1$), we have $|f|^2 \approx \tanh^2 u$ away from the nodal plane ($|f|^2 \approx 1$ on the background). Evaluating $\operatorname{Im}(f^{*}\partial_x f)$ and $\operatorname{Im}(f^{*}\partial_y f)$ yields
\begin{subequations}
\begin{align}
    \operatorname{Im}(f^{*}\partial_x f) &= -\frac{A(t)}{\xi}\sin(ky)\operatorname{sech} u \left[\operatorname{sech}^2 u + \tanh^2 u\right] \notag \\
    & = -\frac{A(t)}{\xi}\operatorname{sech}\left(\frac{x}{\xi}\right)\sin(ky), \\
    \operatorname{Im}(f^{*}\partial_y f) &= k A(t) \tanh\left(\frac{x}{\xi}\right)\operatorname{sech}\left(\frac{x}{\xi}\right)\cos(ky).
\end{align}
\end{subequations}
Retaining the expansion near the soliton core, the velocity field components $\displaystyle v_x = \frac{\operatorname{Im}(f^{*}\partial_x f)}{|f|^2}$ and $\displaystyle v_y = \frac{\operatorname{Im}(f^{*}\partial_y f)}{|f|^2}$ evaluate to
\begin{subequations}
\begin{align}
    v_{x}(x,y,t) &= -\frac{A(t)}{\xi}\left[\operatorname{sech}^{3}\left(\frac{x}{\xi}\right) \right. \notag \\ & \quad \left. + \tanh^{2}\left(\frac{x}{\xi}\right)\operatorname{sech}\left(\frac{x}{\xi}\right)\right]\sin(ky), \\
    v_{y}(x,y,t) &= k A(t)\tanh\left(\frac{x}{\xi}\right)\operatorname{sech}\left(\frac{x}{\xi}\right)\cos(ky).
\end{align}
\end{subequations}

%\bibliography{references}
%\end{document}

\end{document}